\documentclass[twocolumn,aps,prb,reprint,superscriptaddress]{revtex4-1}

\usepackage{graphicx}
\begin{document}
\newcommand{\Eqn}[1]{&\hspace{-0.5em}#1\hspace{-0.5em}&}
\newcommand{\simg}{\stackrel{>}{_\sim}}
\newcommand{\siml}{\stackrel{<}{_\sim}}


\title{Two types of $s$-wave pairing due to magnetic and orbital
fluctuations in the two-dimensional 16-band 
\textit{d}-\textit{p} model for iron-based superconductors}


\author{Yuki Yanagi}
\affiliation{Department of Physics, Niigata University, Ikarashi,
Niigata 950-2181, Japan} 
\author{Youichi Yamakawa}
\affiliation{Department of Physics, Niigata University, Ikarashi,
Niigata 950-2181, Japan} 
\author{Yoshiaki \=Ono}
\affiliation{Department of Physics, Niigata University, Ikarashi,
Niigata 950-2181, Japan} 
\affiliation{Center for Transdisciplinary, Research, Niigata University, Ikarashi, Niigata 950-2181, Japan}


\date{\today}

\begin{abstract}
We study superconductivity in the two-dimensional 16-band $d$-$p$ model extracted from a tight-binding fit to the band structure of LaFeAsO, using the random phase approximation. 
When the intraorbital repulsion $U$ is larger than the interorbital one $U'$, an extended $s$-wave ($s_{\pm}$-wave) pairing with sign reversal of order parameter is mediated by antiferromagnetic spin fluctuations, while when $U<U'$ another kind of $s$-wave ($s_{++}$-wave) pairing without sign reversal is mediated by ferro-orbital fluctuations. 
The $s_{++}$-wave pairing is enhanced due to the electron-phonon coupling and then can be expanded over the realistic parameter region with $U>U'$. 
\end{abstract}


\maketitle

\section{Introduction}
The recently discovered iron-based superconductors\cite{kamihara_1,kamihara}
RFe$Pn$O$_{1-x}$F$_x$ (R=Rare Earth, $Pn$=As, P) with a transition
temperature $T_c$ exceeding 50K\cite{chen_1,ren_1,ren_2,chen_2,ren_3} 
have attracted much
attention. 
The F nondoped compound LaFeAsO exhibits the structural transition from
tetragonal (P4/nmm) to orthorhombic (Cmma) phase 
at a transition temperature $T=$155K and stripe-type antiferromagnetic
order 
at $T=134\mathrm{K}$ with a magnetic moment
$\sim0.36\mu_{B}$\cite{cruz} at low temperature. With increasing F doping, the system becomes metallic and the
antiferromagnetic order disappears\cite{kamihara}, and then, the
superconductivity emerges for $x\sim0.1$ with $T_c\sim26\mathrm{K}$.
 Rare-earth substitution compounds exhibit superconducting transition
 with higher $T_c$\cite{chen_1,ren_1,ren_2,chen_2,ren_3}. 
Specific features of the systems are two-dimensionality of the
conducting Fe$_{2}$As$_{2}$ plane and the orbital degrees of freedom in
Fe$^{2+}$ (3$d^6$)\cite{kamihara_1,kamihara}. 
The pairing symmetry together with the mechanism of 
the superconductivity is one of the most significant issues. 

The NMR Knight shift measurements
revealed that the superconductivity of the systems is the spin-singlet pairing\cite{matano,kawabata_1}. Fully
gapped superconducting states have been predicted by various experiments
such as the penetration depth\cite{hashimoto}, the specific heat\cite{mu}, the angle
resolved photoemission spectroscopy (ARPES)\cite{liu,ding,kondo} and the impurity effect on
$T_c$\cite{kawabata_1,karkin}. 
 In contrast to the above mentioned experiments, the NMR relaxation rate
 shows the power low behavior $1/T{_1}\propto T^3$ below
 $T_{c}$\cite{nakai}, suggesting the nodal or  highly anisotropic gap
 structure. The other NMR measurements\cite{kobayashi}, however, revealed $1/T{_1}\propto T^6$ below
 $T_c$ and there is still controversy.

Theoretically, the first principle calculations have predicted that the
nondoped system is metallic with two or three concentric hole Fermi
surfaces around the $\Gamma$
 point ($\mathbf{k}=(0,0)$) and two elliptical
electron Fermi surfaces around the $M$ 
point ($\mathbf{k}=(\pi,\pi)$)\cite{lebegue,singh,haule,xu,boeri}. Mazin \textit{et al.} suggested
that the spin-singlet extended $s$-wave pairing whose order parameter
changes its sign between the hole pockets and the electron pockets ($s_{\pm}$-wave) is
favored due to the antiferromagnetic spin fluctuations\cite{mazin}.
According to the
weak coupling approaches based on multi-orbital Hubbard models\cite{kuroki_1,kuroki_2,kuroki_3,nomura_2,ikeda,wang_1,graser,yao,stanescu,cvetkovic},
the $s_{\pm}$-wave pairing or the $d_{xy}$-wave pairing is expected to
emerge.  It is shown that the $s_{\pm}$-wave pairing is realized also in the
strong coupling region by the mean field study based on 
the $t$-$J_1$-$J_2$ model\cite{seo} and the exact diagonalization study
based on the one-dimensional two-band
Hubbard model\cite{sano}. 

Generally speaking, the details of the band structure and the Fermi
surface are crucial for determining the pairing symmetry. In the 5-band
Hubbard model originally introduced by Kuroki \textit{et
al.}, the energy
bands obtained by  reproduce those obtained by the density functional
calculation very well\cite{kuroki_1}. In this model, however, the spatial extensions of
the Fe $3d$ like Wannier orbitals are different from each
other\cite{vildosola} and
 the resulting intra-orbital terms of the on-site Coulomb interaction
 are strongly orbital dependent\cite{nakamura}. In addition, since the model explicitly includes the transfer
integrals up to the fifth
nearest neighbor sites\cite{kuroki_1}, one should take the
off-site Coulomb interaction, which is considered to be about
$0.5\mathrm{eV}$ between the nearest
neighbor sites,  into
account to ensure the consistency of the model\cite{nakamura}. On the other hand, in the effective model
which includes both the Fe $3d$ orbitals
and the As $4p$ orbitals, so called $d$-$p$ model, the spatial extensions and the differences of those between the
orbitals will be considerably reduced\cite{vildosola}. Due to these
facts, in the $d$-$p$ model, it is expected that the intra-orbital
terms of the on-site Coulomb interaction for each orbitals have 
almost the same
values and the off-site Coulomb interaction are negligible. Therefore,
theoretical studies based on the $d$-$p$ model, are highly desired. 

 In the previous papers\cite{yamakawa_1,yamakawa_2,yamakawa_4}, we have investigated
the electronic states of the Fe$_2$As$_2$ plane in iron-based
superconductors on the basis of the two-dimensional
16-band $d$-$p$ model which includes the
Coulomb interaction on a Fe site: the intra- and inter-orbital direct
terms $U$ and $U'$, the Hund's coupling $J$ and the pair-transfer
$J'$. Using the random phase
approximation (RPA), we have found that, for a larger value of $J$,
the most favorable pairing symmetry is $s_{\pm}$-wave, while, for a
smaller value of $J$, it is $d_{xy}$-wave.  

The present paper is
 a full paper to our previous papers\cite{yamakawa_1,yamakawa_2,yamakawa_4} with some numerical
improvements\cite{yamakawa_4}. 
In the present
paper, we investigate the superconductivity in the wider parameter
space by treating $U$, $U'$, $J$ and $J'$ as independent parameters in
contrast to the previous study under the condition that
$U=U'+2J$ and $J=J'$ based on the two-dimensional
16-band $d$-$p$ model. Solving the superconducting gap equation with the pairing
interaction obtained by using the RPA, we find that two kinds of the $s$-wave
superconducting states appear. As above mentioned, the $s_{\pm}$-wave superconducting
state emerges near
the incommensurate spin density wave (ISDW) with
$\mathbf{q}\sim(\pi,\pi)$ phase. In addition, for
$U<U'$, the $s$-wave superconducting state appears
around the ferro-orbital
ordered phase. The order parameter for this $s$-wave state dose not change its
sign in $\mathbf{k}$ space. We refer to this $s$-wave state as the
$s_{++}$-wave state, hereafter.
\section{Model and Formulation}
\begin{figure}[t]
\begin{center}
\begin{minipage}{7cm}
\leftline{(a)}
\begin{center}
\includegraphics[width=7cm]{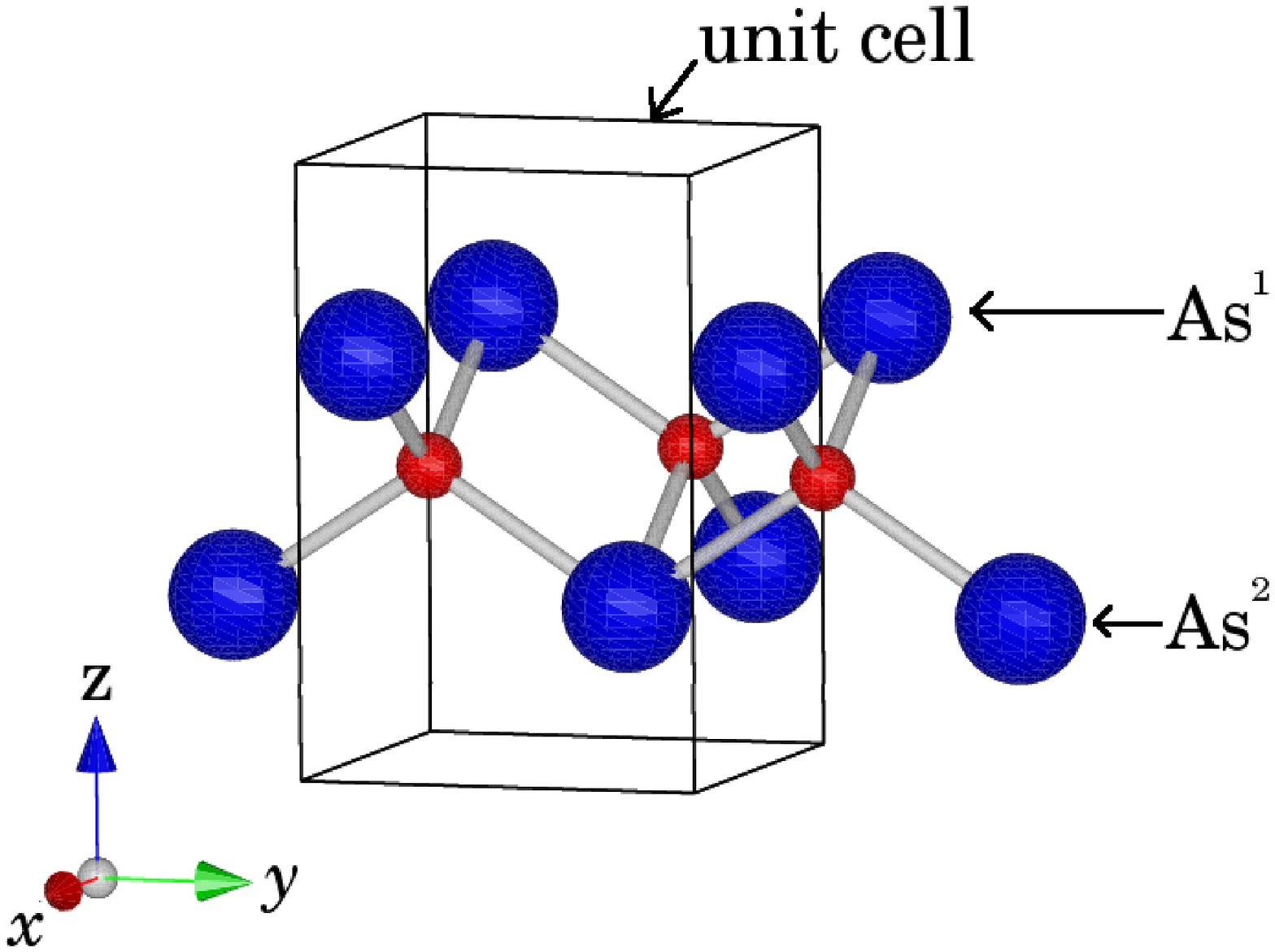}
\end{center}
\end{minipage}
\begin{minipage}{7cm}
\leftline{(b)}
\begin{center}
\includegraphics[width=7cm]{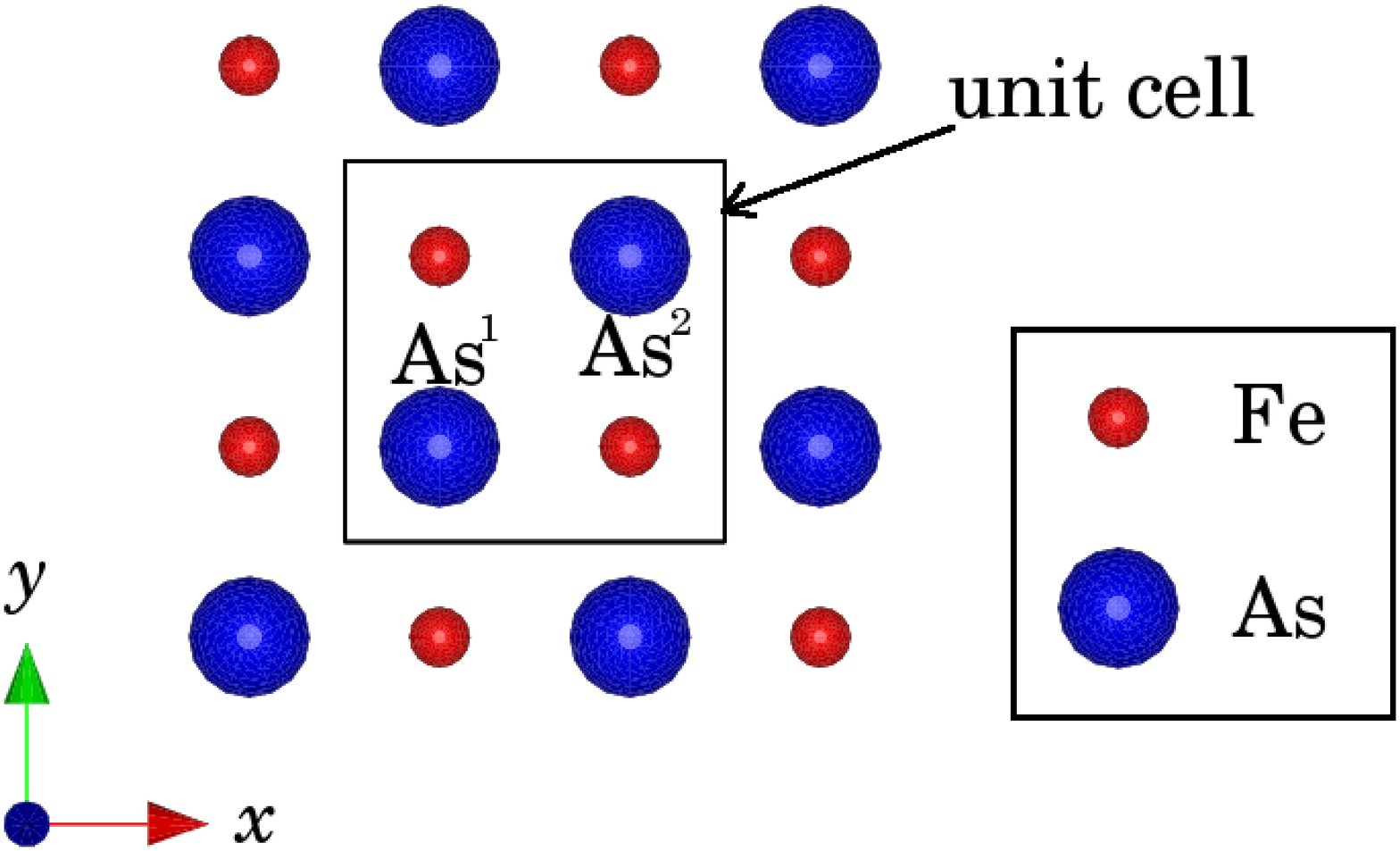}
\end{center}
\end{minipage}
\caption{(Color online) Crystal structure of Fe$_2$As$_2$ layer . Small
 and large balls represent Fe and As atoms, respectively. The solid line
 represents the unit cell. It is noted that As$^1$ and As$^2$ denote the
 As atoms on the upper side and on the lower side of the Fe$_2$As$_2$ layer, respectively\label{crystal}.}
\end{center}
\end{figure}

 First of all, we perform the density functional calculation for LaFeAsO with the
 generalized gradient approximation of Perdew, Burke and Ernzerhof\cite{perdew} by using the WIEN2k
package\cite{blaha}, where the lattice parameters ($a=4.03268$\AA, $c=8.74111$\AA) and
the internal coordinates ($z_{La}=0.14134$, $z_{As}=0.65166$) are
experimentally determined\cite{nomura}. 
 The crystal structure of Fe$_2$As$_2$ layer is shown in Fig. \ref{crystal} (a). 
 Since As atoms are tetrahedrally arranged around a Fe atom, 
there are two distinct Fe and As
sites in the crystallographic unit cell (see Figs. \ref{crystal} (a),
(b)). Considering these facts, we
then derive the two-dimensional 16-band $d$-$p$
model\cite{yamakawa_1,yamakawa_2}, 
where $3d$ orbitals ($d_{3z^2-r^2}$, $d_{x^2-y^2}$, $d_{xy}$, $d_{yz}$, $d_{zx}$) of two Fe
atoms (Fe$^1$=$A$, Fe$^2$=$B$) and $4p$ orbitals ($p_{x}$, $p_{y}$, $p_{z}$) of two As atoms are
explicitly included. We note that
$x, y$ axes are directed along second nearest Fe-Fe
bonds (see Fig. \ref{crystal} (b)). 

The total Hamiltonian of the $d$-$p$ model is given by
\begin{equation}
H=H_0+H_\mathrm{int}, \label{eq_H}
\end{equation}
where $H_0$ and $H_\mathrm{int}$ are the noninteracting and interacting
parts of the Hamiltonian, respectively.
The noninteracting part of the $d$-$p$ model is given by the following tight-binding Hamiltonian,
\begin{eqnarray}
H_0&=&\sum_{i,\ell,\sigma}\hspace{-1mm}
\varepsilon^d_{\ell}d^{\dag}_{i\ell\sigma}d_{i\ell\sigma}
+\hspace{-1mm}\sum_{i,m,\sigma}\hspace{-1mm}\varepsilon^p_{m}p^{\dag}_{im\sigma}p_{im\sigma} \nonumber \\ 
&+&\sum_{i,j,\ell,\ell',\sigma}\hspace{-1mm}t^{dd}_{i,j,\ell,\ell'}d^{\dag}_{i\ell\sigma}d_{j\ell'\sigma}
 +\hspace{-1mm} \sum_{i,j,m,m',\sigma}\hspace{-1mm}t^{pp}_{i,j,m,m'}p^{\dag}_{im\sigma}p_{jm'\sigma} \nonumber\\
&+&\sum_{i,j,\ell,m,\sigma}\hspace{-1mm}t^{dp}_{i,j,\ell,m}d^{\dag}_{i\ell\sigma}p_{jm\sigma}+h.c. \label{d-p}, 
\end{eqnarray}
where $d_{i\ell\sigma}$ is the annihilation operator for Fe-$3d$ electrons with spin
$\sigma$ in the orbital $\ell$ at the site $i$ and $p_{im\sigma}$ is the annihilation
operator for As-$4p$ electrons with spin
$\sigma$ in the orbital $m$ at the site $i$. In eq. (\ref{d-p}), the
transfer integrals $t^{dd}_{i,j,\ell,\ell'}$, $t^{pp}_{i,j,m,m'}$,
$t^{dp}_{i,j,\ell,m}$ and the atomic energies $\varepsilon^d_{\ell}$,
$\varepsilon^p_{m}$ are determined so as to fit both the energy and the
weights 
of orbitals for each band obtained from the
tight-binding approximation to
those from the density functional calculation\cite{yamakawa_3}.  Similar
models have been used by the other authors\cite{cvetkovic,cao,manousakis} but
the model parameters are different from ours. The doping concentration $x$
corresponds to the number of electrons per unit cell $n=24+2x$ in the present
model. 

Now we consider the effect of the Coulomb interaction on Fe site.
The interacting part of the Hamiltonian is given as follows, 
\begin{eqnarray}
H_{\rm int}&=&\frac{1}{2}U\sum_{i}\sum_{\ell}\sum_{\sigma\neq\bar{\sigma}}
d^{\dag}_{i\ell\sigma}d^{\dag}_{i\ell\bar{\sigma}}
d_{i\ell\bar{\sigma}}d_{i\ell\sigma} \nonumber \\
&+&\frac{1}{2}U'\sum_{i}\sum_{\ell\neq\bar{\ell}}\sum_{\sigma,\sigma'}
d^{\dag}_{i\ell\sigma}d^{\dag}_{i\bar{\ell}\sigma'}
d_{i\bar{\ell}\sigma'}d_{i\ell\sigma} \nonumber \\
&+&\frac{1}{2}J\sum_{i}\sum_{\ell\neq\bar{\ell}}\sum_{\sigma,\sigma'}
d^{\dag}_{i\ell\sigma}d^{\dag}_{i\bar{\ell}\sigma'}
d_{i\ell\sigma'}d_{i\bar{\ell}\sigma} \nonumber \\
&+&\frac{1}{2}J'\sum_{i}\sum_{\ell\neq\bar{\ell}}\sum_{\sigma\neq\bar{\sigma}}
d^{\dag}_{i\ell\sigma}d^{\dag}_{i\ell\bar{\sigma}}
d_{i\bar{\ell}\bar{\sigma}}d_{i\bar{\ell}\sigma}, \label{eq_H_int}
\end{eqnarray}
where $U$ and $U'$ are the intra- and inter-orbital direct terms,
respectively, and $J$ and $J'$ are the Hund's coupling and the
pair-transfer, respectively.   
For the isolated atoms, the relations between Coulomb matrix  elements
$U=U'+2J$ and $J=J'$ are derived due to the rotational invariance of  the  Coulomb
interaction and the reality of the wave functions, respectively\cite{tang}. 
 For the atoms in the crystal, however, the
relation is not satisfied generally due to
the crystallographic effects  
 and the many body effects due to the Coulomb interaction and
 the electron-phonon coupling which will be discussed later. 
Therefore, we treat 
$U$, $U'$, $J$ and $J'$ as independent
parameters in the present paper.

Within the RPA\cite{takimoto,mochizuki,yada}, 
the spin susceptibility $\hat{\chi^s}(\mathbf{q})$ and 
the charge-orbital susceptibility $\hat{\chi^c}(\mathbf{q})$ 
are given in the $50\times50$
matrix representation as follows\cite{yamakawa_1,yamakawa_2}, 
\begin{eqnarray}
\hat{\chi^s}(\mathbf{q})&=&(\hat{1}-\hat{\chi}^{(0)}(\mathbf{q})\hat{S})^{-1}\hat{\chi}^{(0)}(\mathbf{q}) \label{eq_chis},\\
\hat{\chi^c}(\mathbf{q})&=&(\hat{1}+\hat{\chi}^{(0)}(\mathbf{q})\hat{C})^{-1}\hat{\chi}^{(0)}(\mathbf{q}) \label{eq_chic}
\end{eqnarray}
with the noninteracting susceptibility
\begin{eqnarray}
&&\chi^{(0)~\alpha,\beta}_{\ell_1\ell_2,\ell_3\ell_4}(\mathbf{q})=
-\frac{1}{N}\sum_{\mathbf{k}}\sum_{\mu,\nu}
\frac{f(\varepsilon_{\mathbf{k}+\mathbf{q},\mu})
-f(\varepsilon_{\mathbf{k},\nu})}
{\varepsilon_{\mathbf{k}+\mathbf{q},\mu}
-\varepsilon_{\mathbf{k},\nu}} \nonumber\\
&\times&u^{\alpha}_{\ell_1,\nu}(\mathbf{k})^*
{u^{\alpha}_{\ell_2,\mu}(\mathbf{k}+\mathbf{q})}
u^{\beta}_{\ell_3,\nu}(\mathbf{k})
{u^{\beta}_{\ell_4,\mu}(\mathbf{k}+\mathbf{q})}^* \label{eq_chi0},
\end{eqnarray}
where $\mu$, $\nu$ (=1-16) are band indexes, $\alpha$, $\beta$ ($=$$A,B$) represent two
Fe sites, $\ell$ represents Fe 3$d$ orbitals,
$u^{\alpha}_{\ell,\mu}(\mathbf{k})$ is the eigenvector which
diagonalizes $H_0$ eq. (\ref{d-p}), $\varepsilon_{\mathbf{k},\mu}$ is
the corresponding eigenenergy of band
$\mu$ with wave vector $\mathbf{k}$ and $f(\varepsilon)$ is the Fermi
distribution function. In eqs. (4) and (5), the interaction matrix $\hat{S}$ ($\hat{C}$) is given by
\begin{equation}
\hat{S}~(\hat{C})= \left\{
\begin{array}{@{\,} l @{\,} c}
U~(U) & (\alpha=\beta,~\ell_1=\ell_2=\ell_3=\ell_4)\\
U'~(-U'+2J) & (\alpha=\beta,~\ell_1=\ell_3\ne\ell_2=\ell_4)\\
J~(2U'-J) & (\alpha=\beta,~\ell_1=\ell_2\ne\ell_3=\ell_4)\\
J'~(J')& (\alpha=\beta,~\ell_1=\ell_4\ne\ell_2=\ell_3)\\
0 & (\mathrm{otherwise})
\end{array} \right. . \label{eq-U}
\end{equation} 

In the weak coupling regime, the superconducting gap equation is given
by\cite{yamakawa_1,yamakawa_2} 
\begin{eqnarray}
&&\lambda\Delta^{\alpha\beta}_{\ell\ell'}({\bf k})=\frac{1}{N}\sum_{{\bf k}'}
\sum_{\ell_1\ell_2\ell_3\ell_4}\sum_{\alpha',\beta'}\sum_{\mu,\nu}\ \ \ \ \ \ \ \ \ \ \ \ \ \ \nonumber\\
&\times&\frac{f(\varepsilon_{-{\bf k}',\mu})
+f(\varepsilon_{{\bf k}',\nu})-1}{\varepsilon_{-{\bf k}',\mu}
+\varepsilon_{{\bf k}',\nu}}V^{\alpha,\beta}_{\ell\ell_1,\ell_2\ell'}
({\bf k}-{\bf
 k}'){\Delta}^{\alpha'\beta'}_{\ell_3\ell_4}({\bf k}')\nonumber\\
&\times&u^{\alpha'}_{\ell_3,\mu}(-{\bf k}'){u^{\alpha}_{\ell_1,\mu}(-{\bf k}')}^*
u^{\beta'}_{\ell_4,\nu}({\bf k}'){u^{\beta}_{\ell_2,\nu}({\bf k}')}^* \label{eq_gap}, \label{gapeq}
\end{eqnarray}
where $\Delta^{\alpha\beta}_{\ell\ell'}({\bf k})$ is the gap function
and $V^{\alpha,\beta}_{\ell_1\ell_2,\ell_3\ell_4}(\mathbf{q})$ is the
effective pairing interaction \cite{notation}.
Within the RPA\cite{takimoto,mochizuki,yada}, $V^{\alpha,\beta}_{\ell_1\ell_2,\ell_3\ell_4}(\mathbf{q})$ 
is given in the $50\times50$
matrix,
\begin{equation}
\hat{V}(\mathbf{q})=\eta\left(\hat{S}\hat{\chi}^s(\mathbf{q})\hat{S}
+\frac{1}{2}\hat{S}\right)-\frac{1}{2}\left(\hat{C}\hat{\chi}^c(\mathbf{q})\hat{C}-\frac{1}{2}\hat{C}\right)\label{eq_veff_s},
\end{equation}
where $\eta=\frac{3}{2}$ for the spin-singlet state and
$\eta=-\frac{1}{2}$  for the spin-triplet state.
The gap equation (\ref{eq_gap}) is solved to
obtain the gap function $\Delta^{\alpha\beta}_{\ell\ell'}(\mathbf{k})$
with the eigenvalue $\lambda$. At $T=T_c$, the largest eigenvalue $\lambda$
becomes unity.
In the present paper, we only focus on the case with $x=0.1$, where
the superconductivity is observed in the compounds\cite{kamihara}.
For simplicity, we set $x=0.1$ and $T=0.02\mathrm{eV}$ in the present study. We use $32\times32$
$\mathbf{k}$ points in the numerical calculations for eqs. (\ref{eq_chis})-(\ref{eq_veff_s}), and also use the fast Fourier
transformation (FFT) to
solve the gap equation eq. (\ref{eq_gap}). Here and hereafter, we
measure the energy in units of eV.
\begin{table}[t]
\begin{center}
\begin{tabular}{ccc} \hline
 on-site energy & & \\ \hline
 $d_{3z^2-r^2}$ & -0.687 & \\ 
 $d_{x^2-y^2}$ & -0.610 & \\ 
 $d_{xy}$ &  -0.921 & \\ 
 $d_{yz}$ & -0.820 & \\ 
 $p_{x}$ & -1.789 & \\ 
 $p_{z}$ & -2.173 & \\ \hline
 $d$-$d$ hopping & nearest & next nearest \\ \hline 
$d_{3z^2-r^2}$-$d_{3z^2-r^2}$ & -0.008 & -0.024 \\
$d_{x^2-y^2}$-$d_{x^2-y^2}$ & 0.143 & -0.023 \\
$d_{xy}$-$d_{xy}$ & 0.328 & 0.073 \\
$d_{yz}$-$d_{yz}$ & 0.109 & -0.012 \\
$d_{zx}$-$d_{zx}$ & 0.109 & 0.012 \\
$d_{3z^2-r^2}$-$d_{xy}$ & 0.078 & \\
$d_{3z^2-r^2}$-$d_{x^2-y^2}$ &  & -0.184 \\ 
$d_{yz}$-$d_{zx}$ & 0.184  &  \\ \hline
 $p$-$p$ hopping & As$^1$-As$^1$ & As$^1$-As$^2$ \\ \hline
$p_{x}$-$p_{x}$  & 0.650  & 0.311 \\ 
$p_{y}$-$p_{y}$  & 0.027  & 0.311  \\ 
$p_{z}$-$p_{z}$ & 0.048  &  0.389 \\ 
$p_{x}$-$p_{y}$ &  & 0.111  \\ 
$p_{x}$-$p_{z}$ &  & 0.297  \\ \hline
$d$-$p$ hopping & nearest & \\ \hline
$d_{3z^2-r^2}$-$p_x$ & 0.646 & \\
$d_{3z^2-r^2}$-$p_z$ & -0.291 & \\
$d_{x^2-y^2}$-$p_x$ & 0.276 & \\
$d_{x^2-y^2}$-$p_z$ & 0.563 & \\
$d_{xy}$-$p_y$ & 0.694 &  \\
$d_{yz}$-$p_y$ & 0.319 & \\
$d_{zx}$-$p_x$ & 0.783 & \\
$d_{zx}$-$p_z$ & 0.164 & 
\end{tabular}
\caption{Tight-binding parameters (in units of eV) for the $d$-$p$ Hamiltonian
 eq. (\ref{d-p}). It is noted that we define the $d$-$p$ hopping and the in-plane $p$-$p$ hopping
 parameters along $x$-axis. \label{table-d-p}}
\end{center}
\end{table}
\begin{figure}[t]
\begin{center}
\begin{minipage}{40mm}
\begin{center}
\includegraphics[width=40mm]{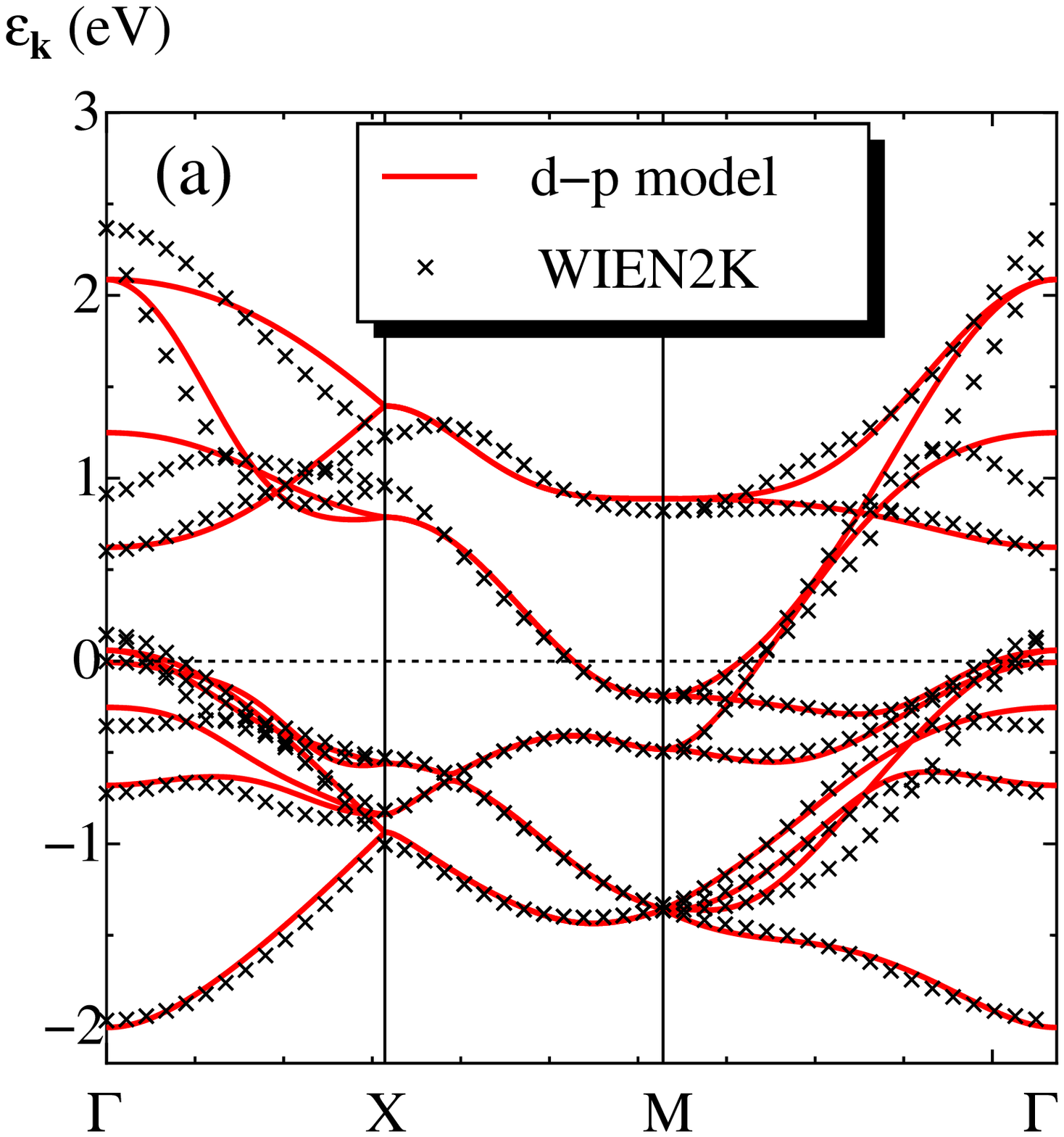}
\end{center}
\end{minipage}
\begin{minipage}{37mm}
\begin{center}
\includegraphics[width=37mm]{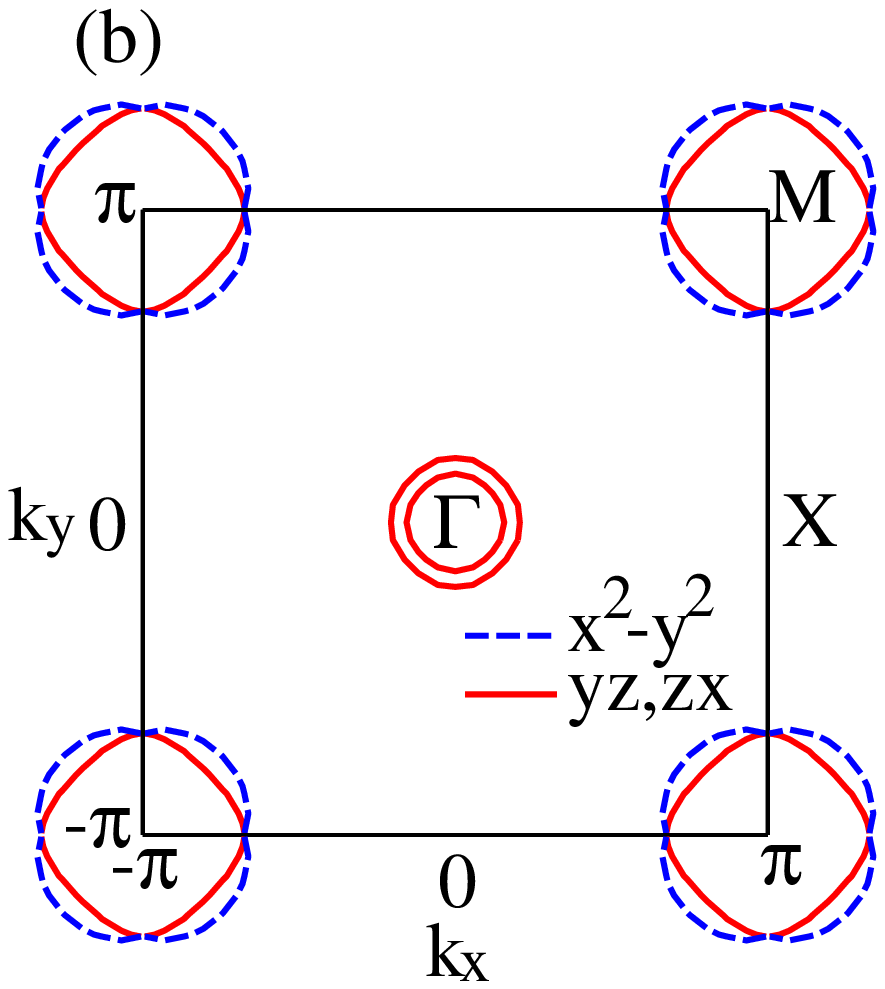}
\end{center}
\end{minipage}
\hspace{2mm}
\begin{minipage}{55mm}
\begin{center}
\includegraphics[width=55mm]{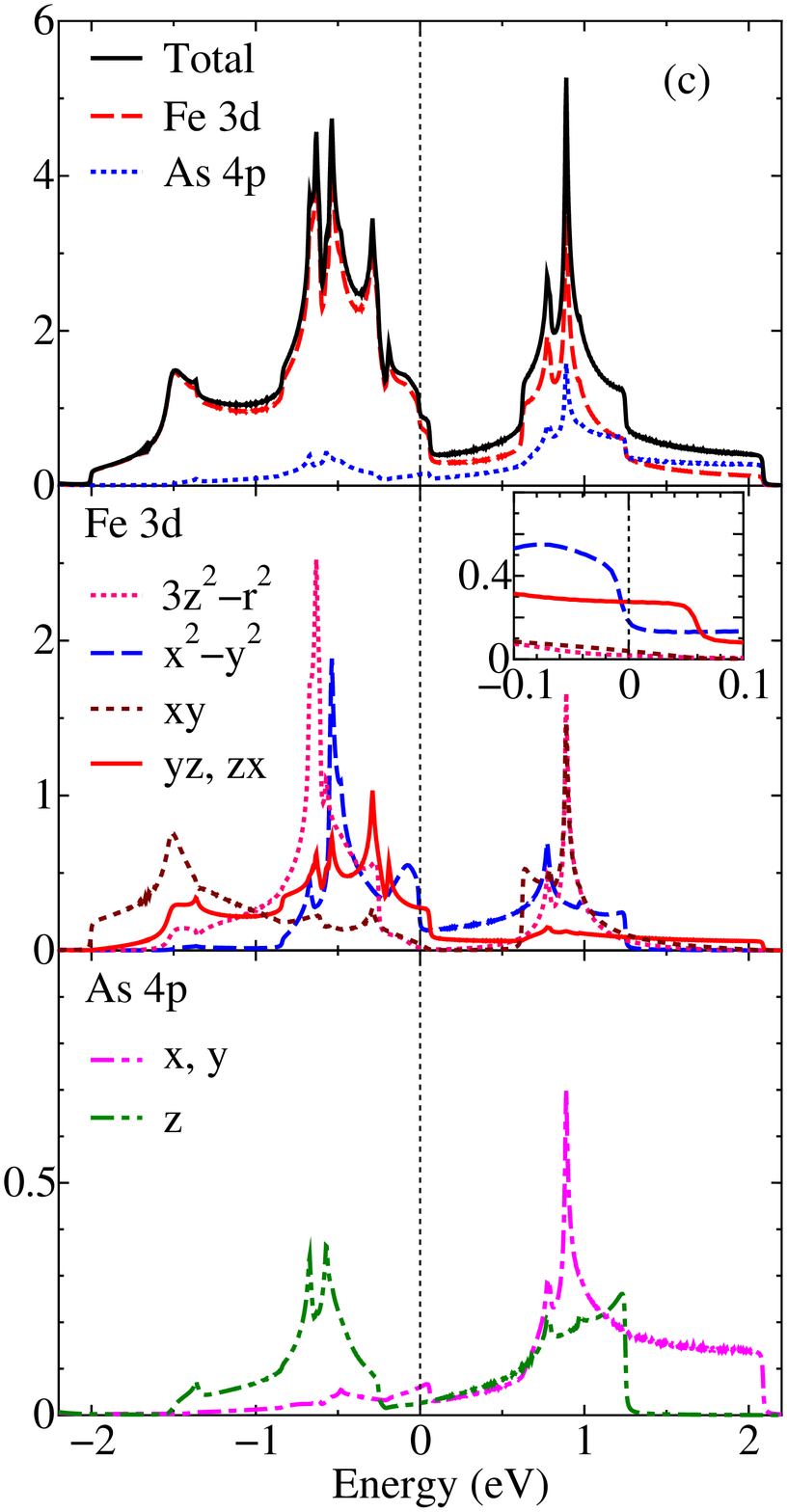}
\end{center}
\end{minipage}
\end{center}
\vspace{-3.5mm}
\caption{(Color online) (a) The band structure obtained from the $d$-$p$
 model eq. (\ref{d-p})
 (solid line)
 and that obtained from the density functional calculation (cross) for $x=0.1$. 
(b) Fermi surface obtained from the $d$-$p$ model for
 $x=0.1$. The solid and dashed lines show the Fermi surfaces which have
 mainly $d_{yz}, d_{zx}$ and $d_{x^2-y^2}$ orbital character, respectively.
(c) The DOS obtained from the $d$-$p$ model for $x=0.1$. Upper panel: total DOS, Middle panel:
 partial DOS of the Fe $3d$ orbitals, Lower panel: partial DOS of the As $4p$
 orbitals. The inset of the middle panel shows the DOS near the Fermi
 level. We note that the Fermi level set to 0 on the energy axis. \label{FS}}
\end{figure}
\section{Calculated Results}
\subsection{Band Structure}
We show the band structure obtained from the $d$-$p$ tight-binding
Hamiltonian eq. (\ref{d-p}), where the
tight-binding parameters are listed in Table \ref{table-d-p}\cite{yamakawa_3}, together with that 
obtained from 
the density functional calculation in the Fig. \ref{FS} (a).  
The result of our density functional calculation  
is similar to that previously 
reported by the other
authors\cite{lebegue,singh,haule,xu,boeri,kuroki_1}.  
It is found that the former reproduces the latter very 
well. We note that the weights of orbitals also agree very well with
each other 
(not shown). Due to the weak crystalline electric field from the As$^{3-}$ ions tetrahedrally
arranged around a Fe atom and the strong hybridization between the Fe
$3d$ orbitals, the
 resulting energy bands have very complicated structure.

The Fermi surface
for the $d$-$p$ tight-binding Hamiltonian is shown in Fig. \ref{FS} (b), where
we can see nearly circular hole pockets around the $\Gamma$
point  and elliptical electron pockets around the
$M$ point. These results are consistent with
the previous first principle calculations\cite{lebegue,singh,haule,xu,boeri}. 

 The density of states (DOS) obtained by the $d$-$p$ tight-binding
 Hamiltonian 
 eq. (\ref{d-p}) is shown in Fig. \ref{FS} (c). It is found that the dominant
 contribution near the Fermi level comes from Fe $3d$ orbitals and the  
 contribution of As $4p$ orbitals is small but is not negligible. We
 show the partial DOS of Fe $3d$ orbitals and that of As $4p$ orbitals in
 the middle panel and the lower panel of Fig. \ref{FS} (c), respectively. The $d_{yz}$,
 $d_{zx}$ and $d_{x^2-y^2}$ states comprise the large part of the DOS
 near the Fermi level, while, the $d_{3z^2-r^2}$, $d_{xy}$ states occupy the
 small one and are comparable with the $p_{x}$, $p_y$ and $p_z$
 states. The $d_{yz}$ and $d_{zx}$ states at the Fermi level are larger
 than the $d_{x^2-y^2}$ ones and this corresponds to the fact that the electron
 pockets have  $d_{yz}$, $d_{zx}$ and $d_{x^2-y^2}$ orbital characters,
 while, the hole pockets have only $d_{yz}$ and $d_{zx}$ orbital
 characters. However, the $d_{x^2-y^2}$ states have large values just
 below the Fermi level as shown in the inset of the middle panel of
 Fig. \ref{FS} (c). This is due to the hole band near the $\Gamma$-point just
 below the Fermi level.
Therefore, it is anticipated that the $d_{yz}$, $d_{zx}$ and
 $d_{x^2-y^2}$ orbitals play significant roles to determine the magnetic,
 orbital and superconducting properties. 

\subsection{RPA Results for $U>U'$}
\begin{figure*}[t]
\begin{center}
\includegraphics[width=40.0mm]{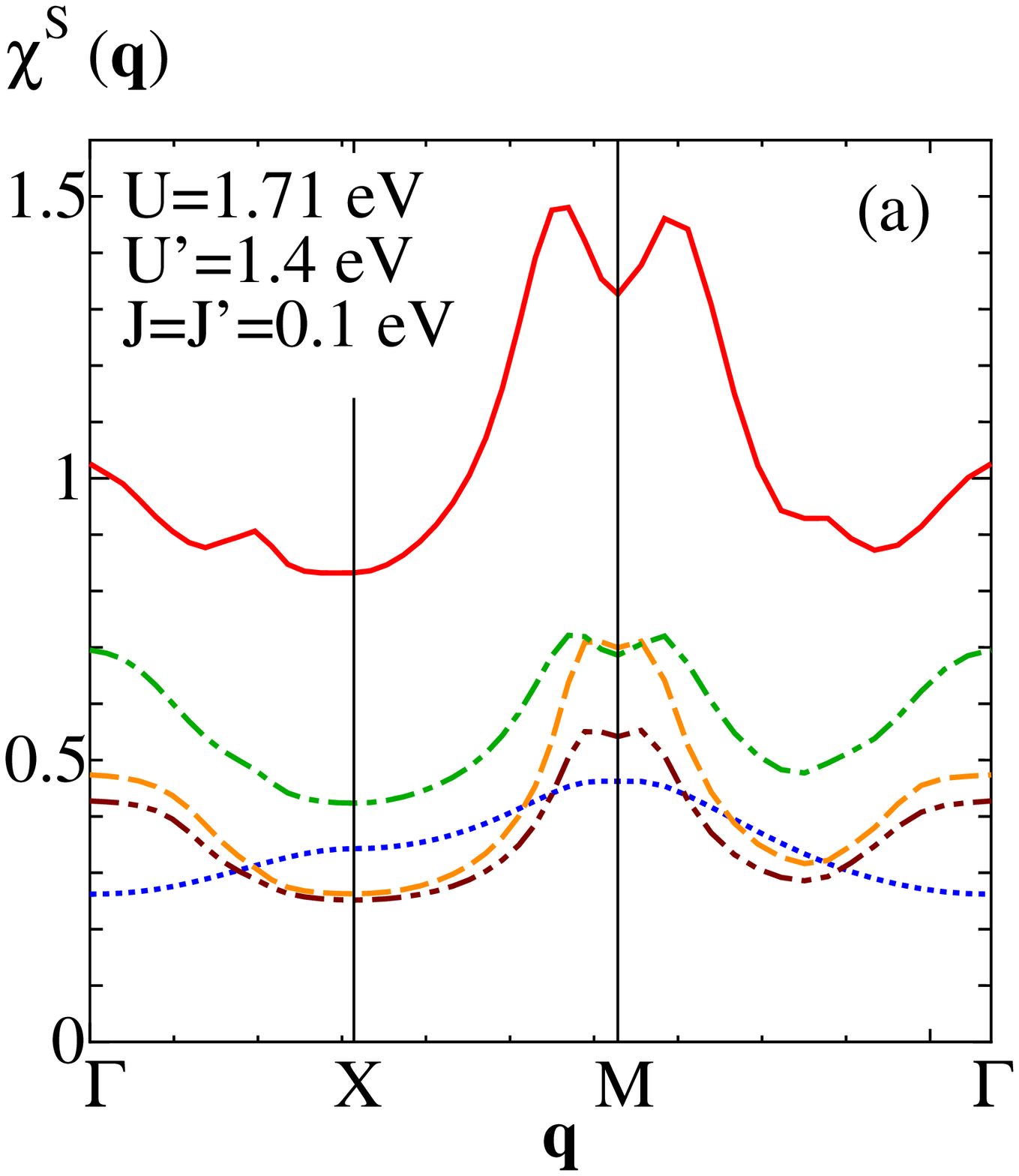}
\includegraphics[width=40.0mm]{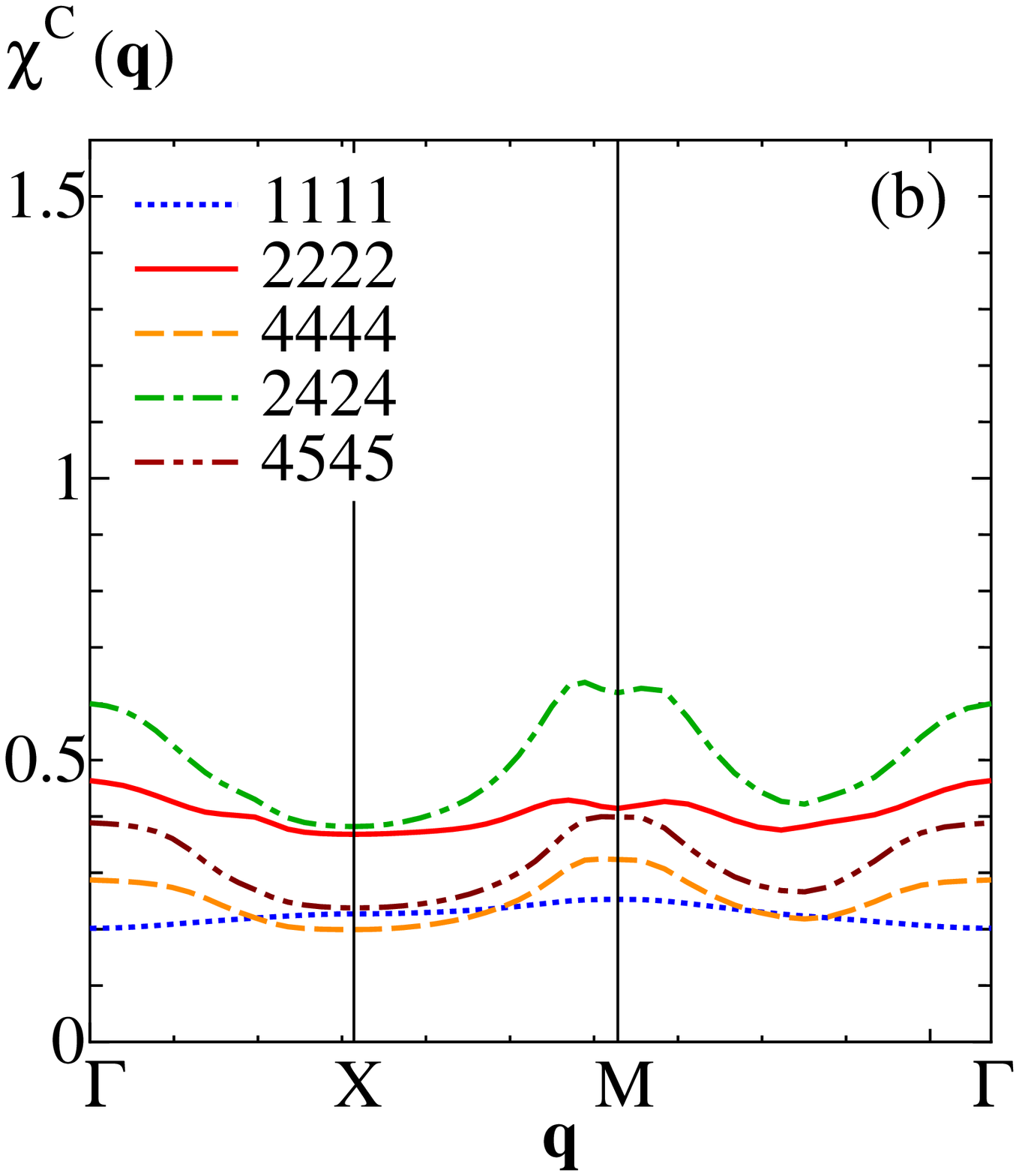}
\includegraphics[width=40.0mm]{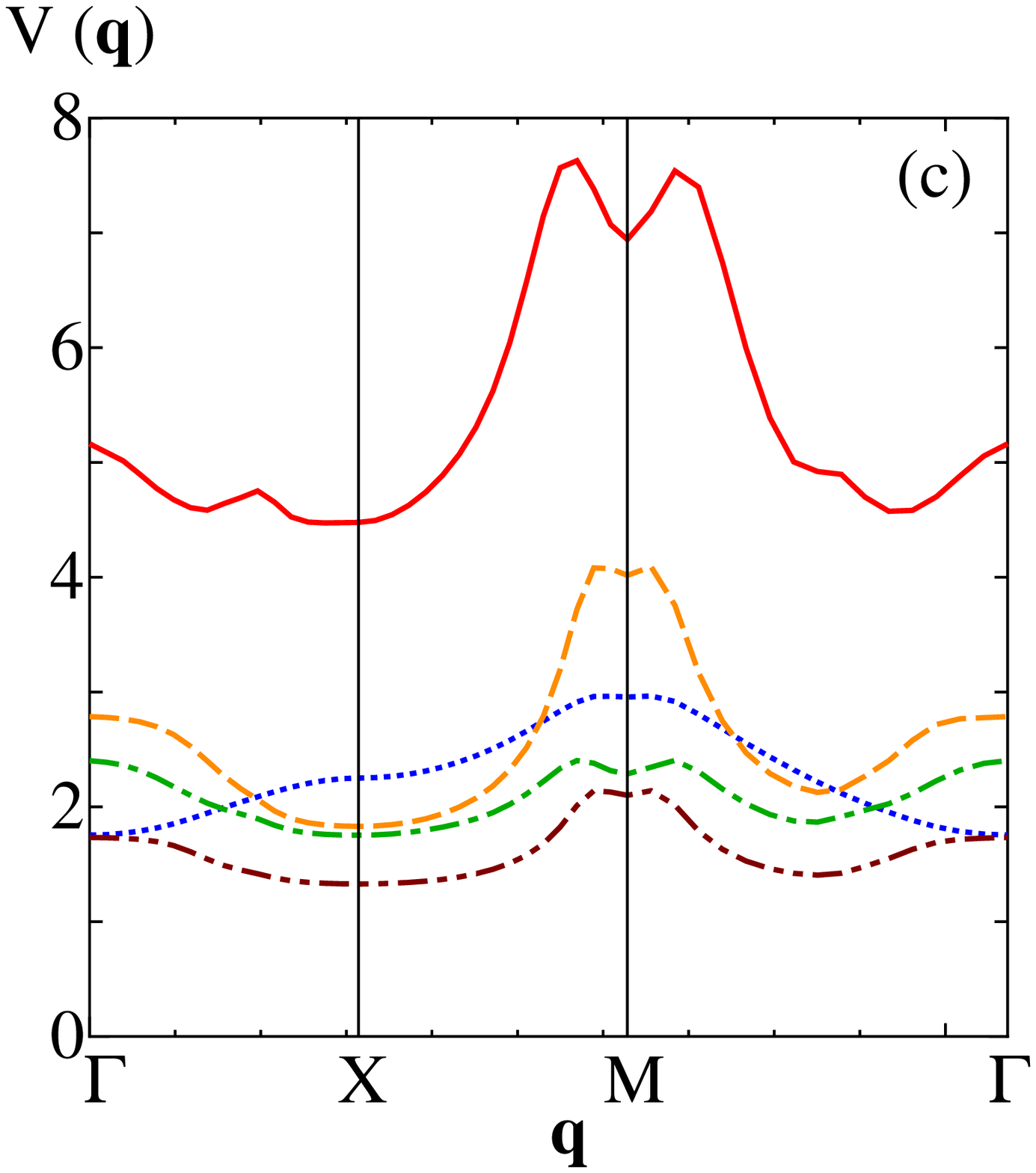}
\includegraphics[width=40.0mm]{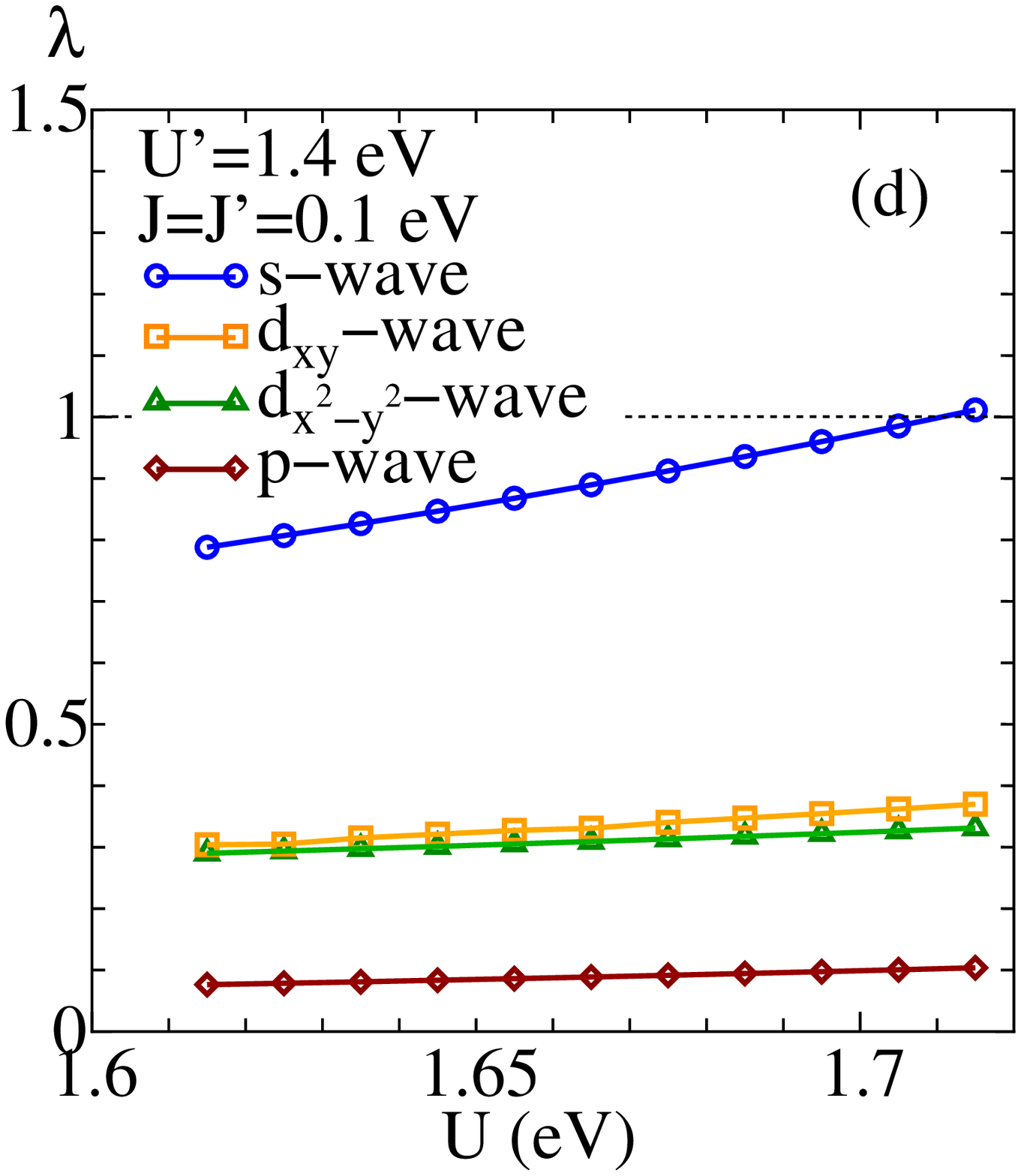}
\caption{(Color online) (a) Several components of the spin
 susceptibility $\hat{\chi}^s(\mathbf{q})$,
  (b) the charge-orbital susceptibility $\hat{\chi}^c(\mathbf{q})$ and (c)
  the effective pairing interaction $\hat{V}(\mathbf{q})$  in $\mathbf{q}$
 space for $U=1.71$, $U'=1.4$ and $J=J'=0.1\mathrm{eV}$. It is noted that we number the orbitals as follows:
 $d_{3z^2-r^2}$(1), $d_{x^2-y^2}$(2), $d_{xy}$(3), $d_{yz}$(4),
 $d_{zx}$(5). (d) $U$ dependence of the eigenvalues of the gap equation
 for several symmetries. \label{chi}}
\end{center}
\end{figure*}

 In this subsection, we concentrate our attention on the case with $U>U'$. We set
 the typical parameters as $U=1.71$, $U'=1.4$ and
$J=J'=0.1 $, where the condition for the superconducting
transition $\lambda= 1$  is satisfied  as mentioned below
 (see Fig. \ref{chi} (d)).
\begin{figure*}[t]
\begin{center}
\includegraphics[width=40.0mm]{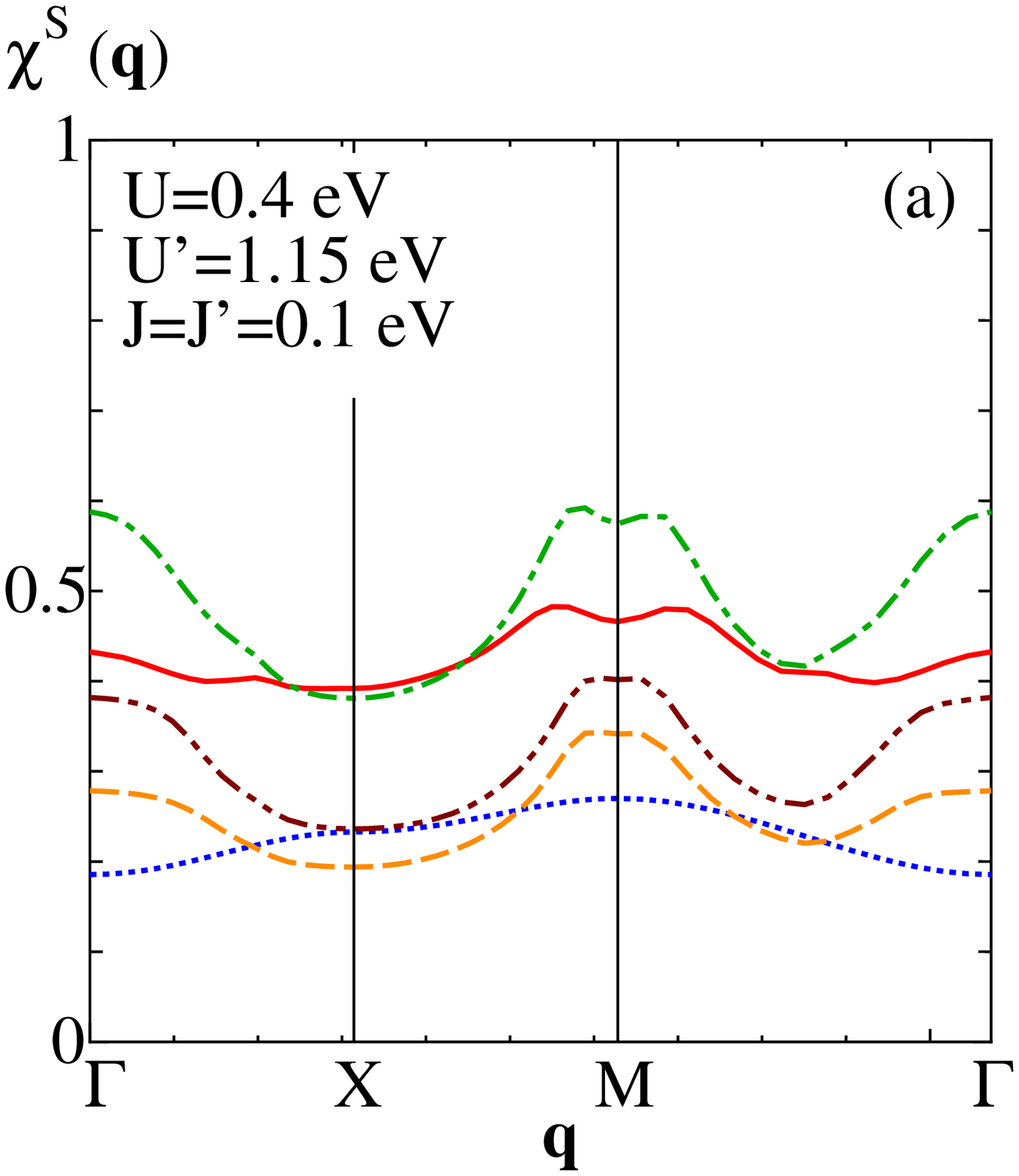}
\includegraphics[width=40.0mm]{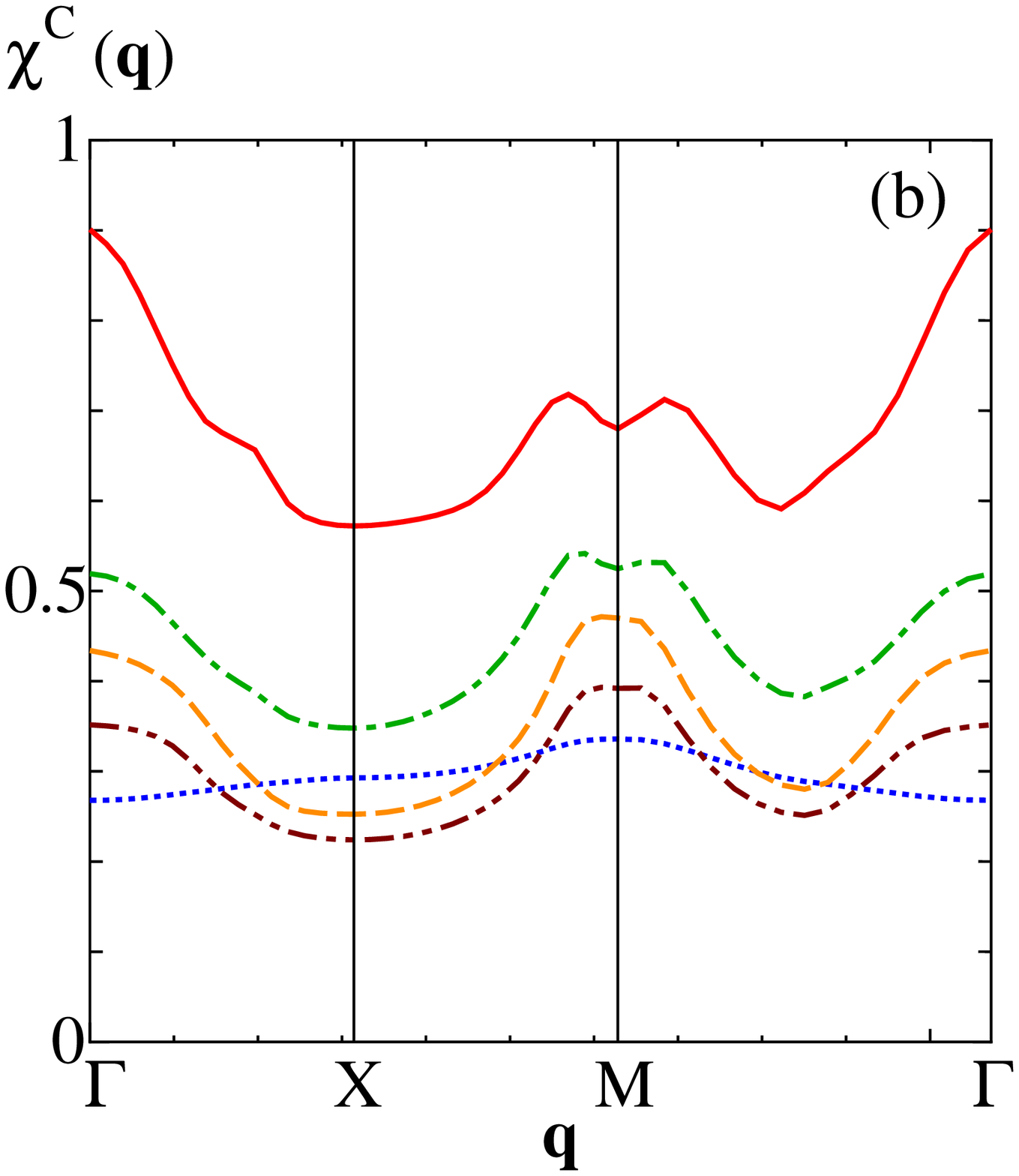}
\includegraphics[width=40.0mm]{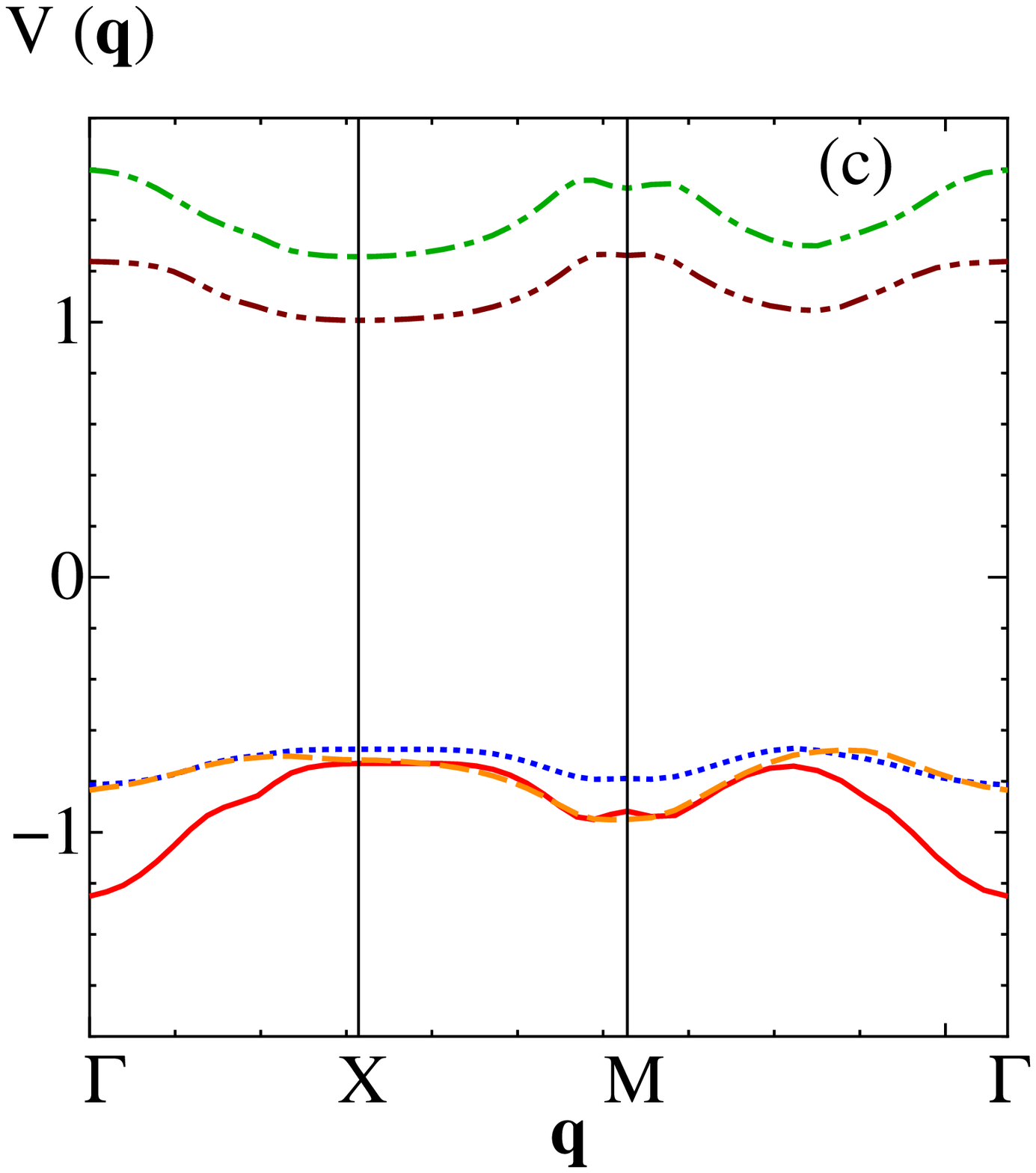}
\includegraphics[width=40.0mm]{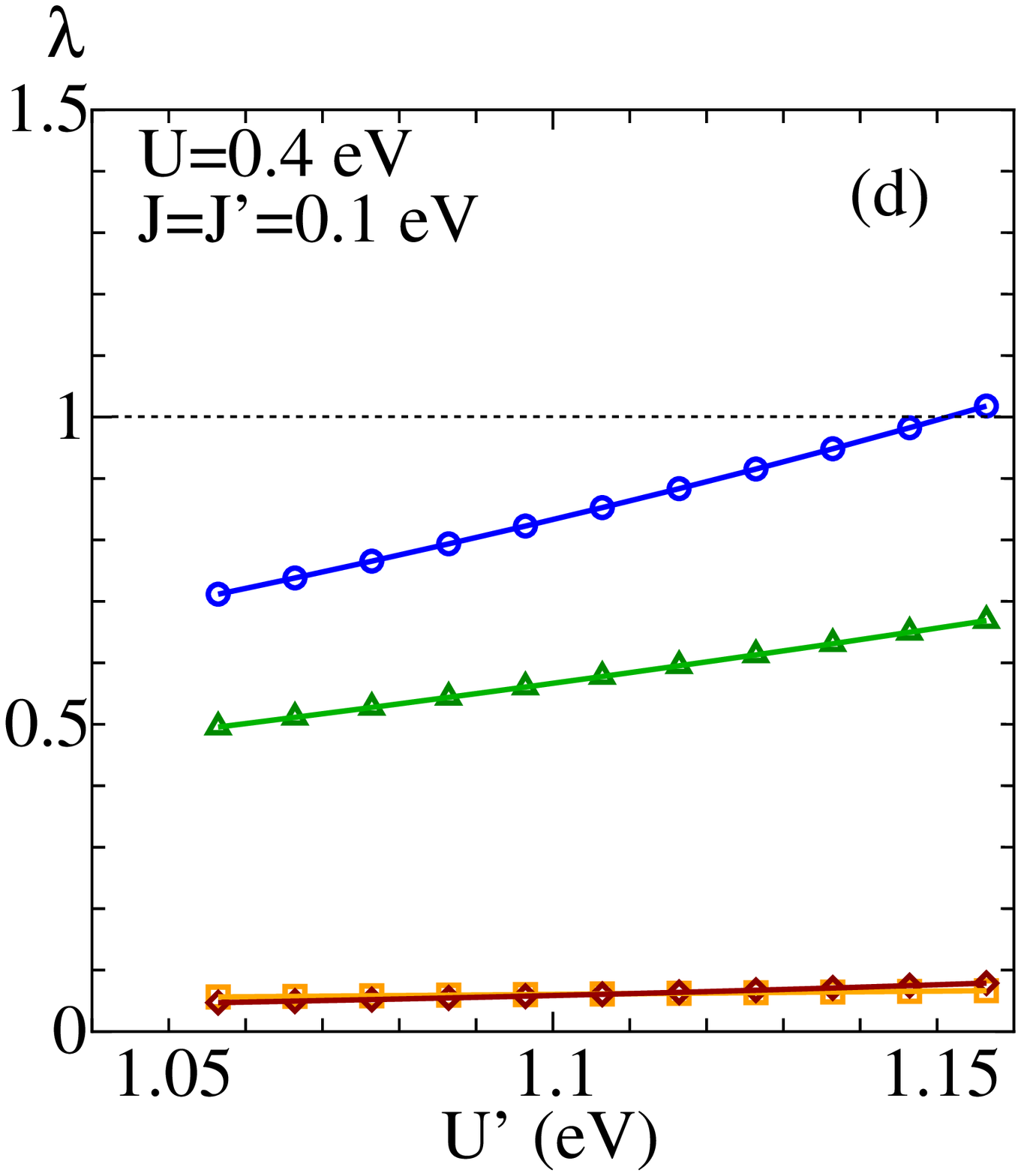}
\caption{(Color online) (a) Several components of the spin
 susceptibility $\hat{\chi}^s(\mathbf{q})$,
  (b) the charge-orbital susceptibility $\hat{\chi}^c(\mathbf{q})$ and (c)
  the effective pairing interaction $\hat{V}(\mathbf{q})$  in $\mathbf{q}$
 space for $U=0.4$, $U'=1.15$ and
 $J=J'=0.1\mathrm{eV}$. It is noted that (d) $U'$ dependence of the eigenvalues of the gap equation
  for several symmetries. It is noted that the legends are the same as
  Fig. \ref{chi} \label{chi-2}}
\end{center}
\end{figure*}

The several components of the spin susceptibility $\hat{\chi}^s (\mathbf{q})$ given in eq. (\ref{eq_chis}) are plotted in Fig. \ref{chi} (a).  The spin susceptibility is enhanced due to the effect of the Coulomb interaction. 
It is found that the most dominant component is the $d_{x^2-y^2}$ diagonal component and the incommensurate peaks around the $M$ point are observed as reflecting the nesting between the hole pockets and the electron pockets. 
As mentioned before, the hole band which has mainly $d_{x^2-y^2}$ orbital character exists just below ($\sim 0.01$eV) the Fermi level and contributes to the large value of the DOS (see Fig. \ref{FS} (c). Therefore, the $d_{x^2-y^2}$ diagonal component of $\hat{\chi}^s (\mathbf{q})$ becomes most dominant at finite temperature $T=0.02$eV ($>0.01$eV). The result is consistent with the RPA results based on the 5-band Hubbard model \cite{kuroki_1}.

The several components of the charge-orbital susceptibility
$\hat{\chi}^c(\mathbf{q})$ given in
eq. (\ref{eq_chic}) are
plotted in Fig. \ref{chi} (b). In contrast to the case with
the spin susceptibility, the off-diagonal component of
$d_{x^2-y^2}-d_{yz}$ which corresponds to the transverse orbital susceptibility 
becomes most dominant 
and  shows
peaks
around the $M$ point together with those at the $\Gamma$ point. It is noted that for
$U>U'$ the spin fluctuations dominate over the charge-orbital
fluctuations as shown in Figs. \ref{chi} (a) and (b).

The several components of the effective pairing interaction
$\hat{V} (\mathbf{q})$ for the spin-singlet state given in eq. (\ref{eq_veff_s}) are
plotted in Fig. \ref{chi} (c). Since the largest eigenvalue $\lambda$ is always
spin-singlet state in the present study, we show the effective pairing
interaction only for the spin-singlet state. Since in the case for $U=1.71$, $U'=1.4$ and
$J=J'=0.1 $, the spin fluctuations  dominate over the orbital
fluctuations as mentioned above, the structures of $\hat{V} (\mathbf{q})$ are
similar to those of the spin susceptibility. 

Substituting $\hat{V} (\mathbf{q})$ into the gap
equation eq. (\ref{eq_gap}), we obtain the gap function
$\hat{\Delta} (\mathbf{k})$ with the eigenvalue
$\lambda$.  In Fig. \ref{chi} (d),
 the eigenvalues $\lambda$ for
various pairing symmetries are plotted as functions of
$U$ for fixed values of $U'$, $J$, $J'$. 
 With increasing $U$, $\lambda$ monotonically increases and 
finally becomes unity at a critical value $U_c$ above which the
superconducting state is realized.
 For  $U'=1.4$ and $J=J'=0.1$
the largest eigenvalue $\lambda$ is for the $s$-wave symmetry and
$U_c=1.71$. The second largest eigenvalue is for $d_{xy}$-wave symmetry
and the eigenvalue for the $d_{xy}$-wave symmetry increases as $J$ increases for $U>U'$. 
 
\subsection{RPA Results for $U<U'$}
 In this subsection, we concentrate our attention on the case with $U<U'$. We set
 the typical parameters as  $U=0.4$, $U'=1.15$ and
$J=J'=0.1$, where the condition for the superconducting
transition $\lambda= 1$  is satisfied as mentioned below
(see Fig. \ref{chi-2} (d)).

 The several components of the spin susceptibility 
$\hat{\chi}^s (\mathbf{q})$ given in
eq. (\ref{eq_chis}) are
plotted in Fig. \ref{chi-2} (a).  In contrast to the case with $U>U'$
(see Fig. \ref{chi} (a)), the off-diagonal
element $d_{x^2-y^2}-d_{yz}$ is most dominant owing to the inter-orbital
direct term $U' > U$.

\begin{figure}[t]
\begin{center}
\includegraphics[width=75mm]{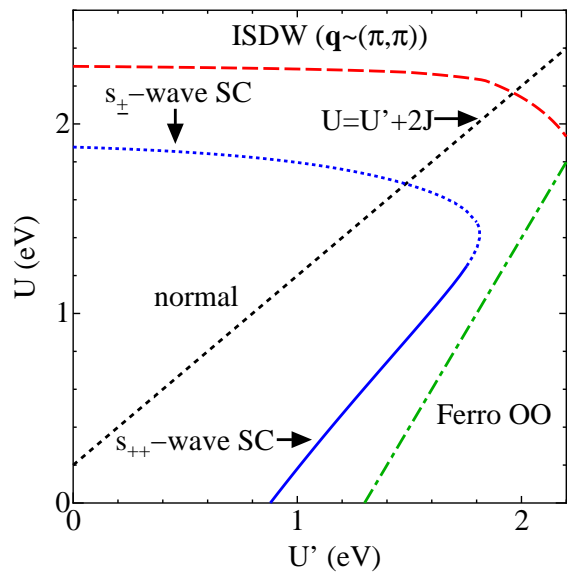}
\caption{(Color online) The phase diagram on $U'$-$U$ plane for
 $J=J'=0.1\mathrm{eV}$ at
 $x=0.1$, $T=0.02\mathrm{eV}$. The solid and dotted lines
 show the $s_{++}$-wave and the $s_{\pm}$-wave superconducting 
 instabilities, respectively. The dashed and dot-dashed lines show instabilities towards the incommensurate spin density wave and the ferro-orbital order, respectively. \label{phasediagram}}
\end{center}
\end{figure}

The several components of the charge-orbital susceptibility
$\hat{\chi}^c(\mathbf{q})$ given in
eq. (\ref{eq_chic}) are
plotted in Fig. \ref{chi-2} (b). In contrast to the case with
the spin susceptibility, the diagonal component of
$d_{x^2-y^2}$ 
becomes most dominant 
and  shows
peaks
around the $\Gamma$ point.  It is noted that for
$U<U'$ the  charge-orbital fluctuations, which corresponds to the
fluctuations near the ferro-orbital ordered state realized in the large
$U'$ regime as mentioned later (see Fig. \ref{phasediagram}), dominate over the spin
fluctuations as shown in Figs. \ref{chi-2} (a) and (b). 

The several components of the effective pairing interaction
$\hat{V} (\mathbf{q})$ for the spin-singlet state given in eq. (\ref{eq_veff_s}) are
plotted in Fig. \ref{chi-2} (c).  Since for
$U=0.4$, $U'=1.15$ and $J=J'=0.1 $, the charge-orbital
fluctuations are larger than the spin fluctuations, the
diagonal components of $\hat{V} (\mathbf{q})$ are always negative in
$\mathbf{q}$ space. 

Substituting $\hat{V} (\mathbf{q})$ into the gap
equation eq. (\ref{eq_gap}), we obtain the gap function
$\hat{\Delta} (\mathbf{k})$ with the eigenvalue
$\lambda$.  In Fig. \ref{chi-2} (d),
 the eigenvalues $\lambda$ for
various pairing symmetries are plotted as functions of
$U'$ for fixed values of $U$, $J$ and $J'$. 
 With increasing $U'$, $\lambda$ monotonically increases and 
finally becomes unity at a critical value $U'_c=1.15$ above which the
superconducting state is realized.
Similar to the case of $U>U'$,  the largest eigenvalue $\lambda$ is for the $s$-wave
symmetry  but the superconducting gap structure is significantly different
from that for $U>U'$ as shown below.

\begin{figure}[t]
\begin{center}
\includegraphics[height=82.0mm,angle=-90]{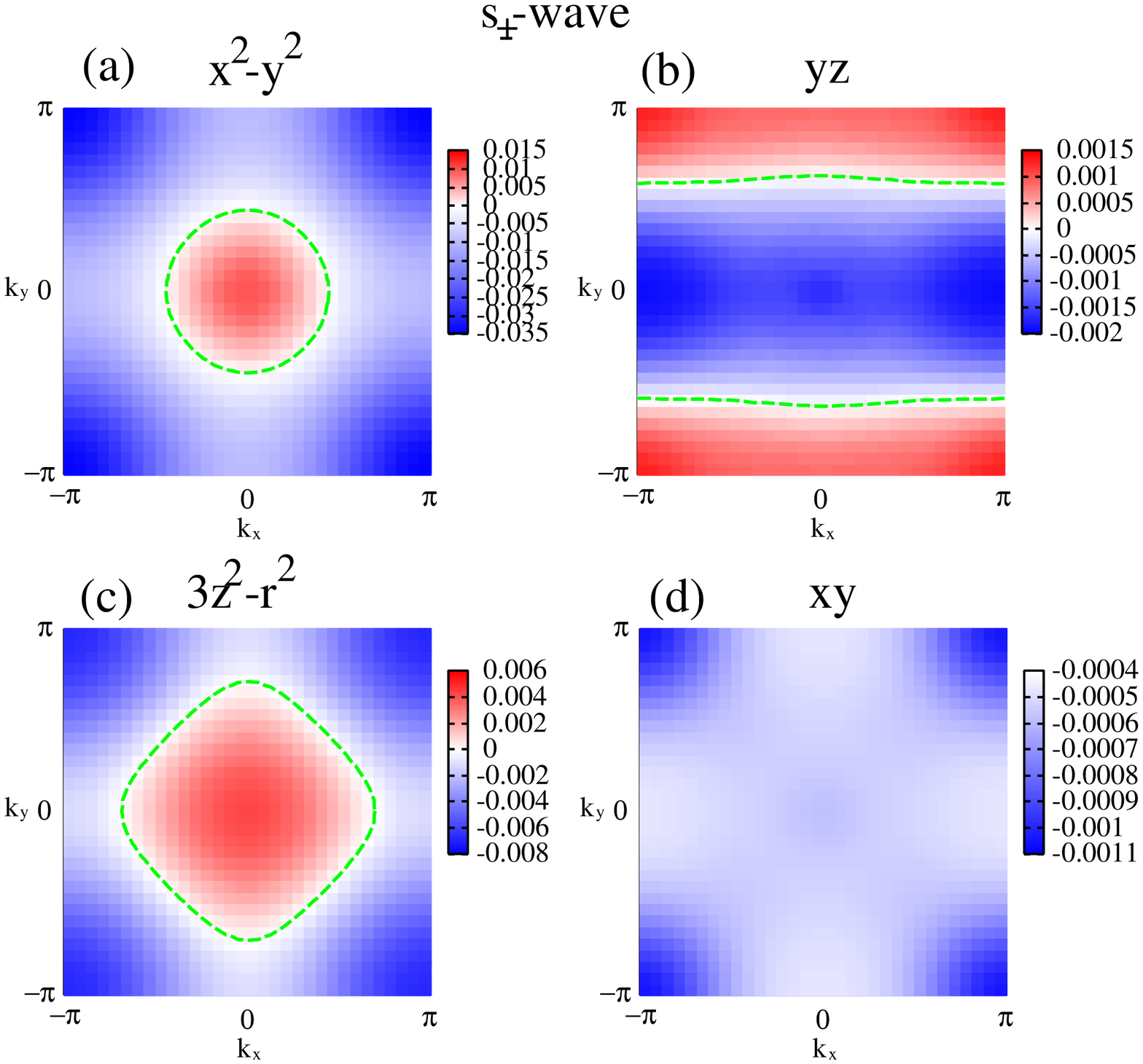}
\includegraphics[height=80.0mm,angle=-90]{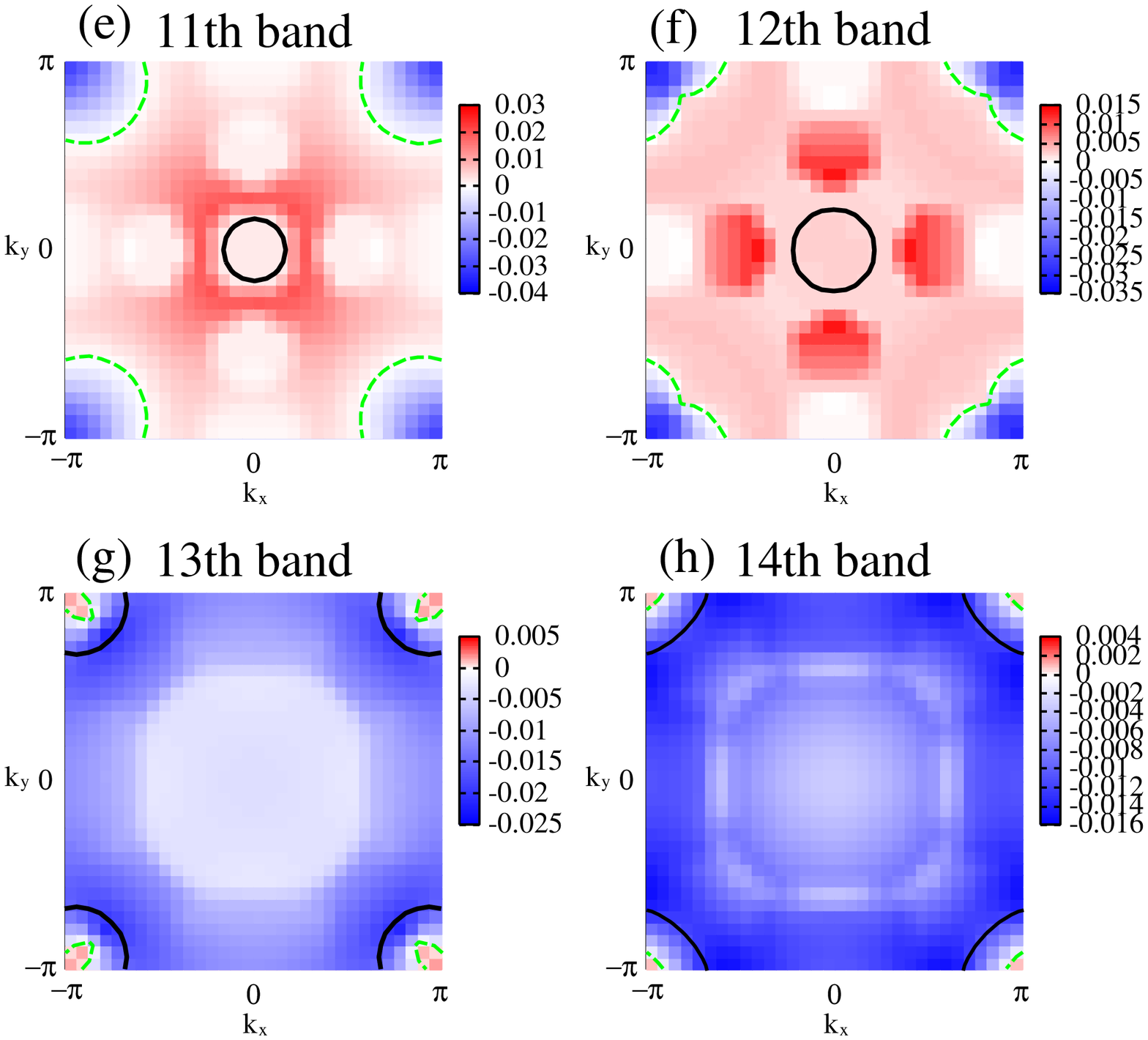}
\caption{(Color online) (a), (b), (c), (d) The diagonal components of the gap function
 $\hat{\Delta} (\mathbf{k})$ in the orbital representation and (e), (f),
 (g), (h) those in
 the band representation for $U=1.71$, $U'=1.4$ and $J=J'=0.1\mathrm{eV}$. The solid
 and dashed lines represent the Fermi
 surfaces and the nodes of the gap function, respectively. \label{fig_gap_1}}
\end{center}
\end{figure}
\begin{figure}[t]
\begin{center}
\includegraphics[height=82.0mm,angle=-90]{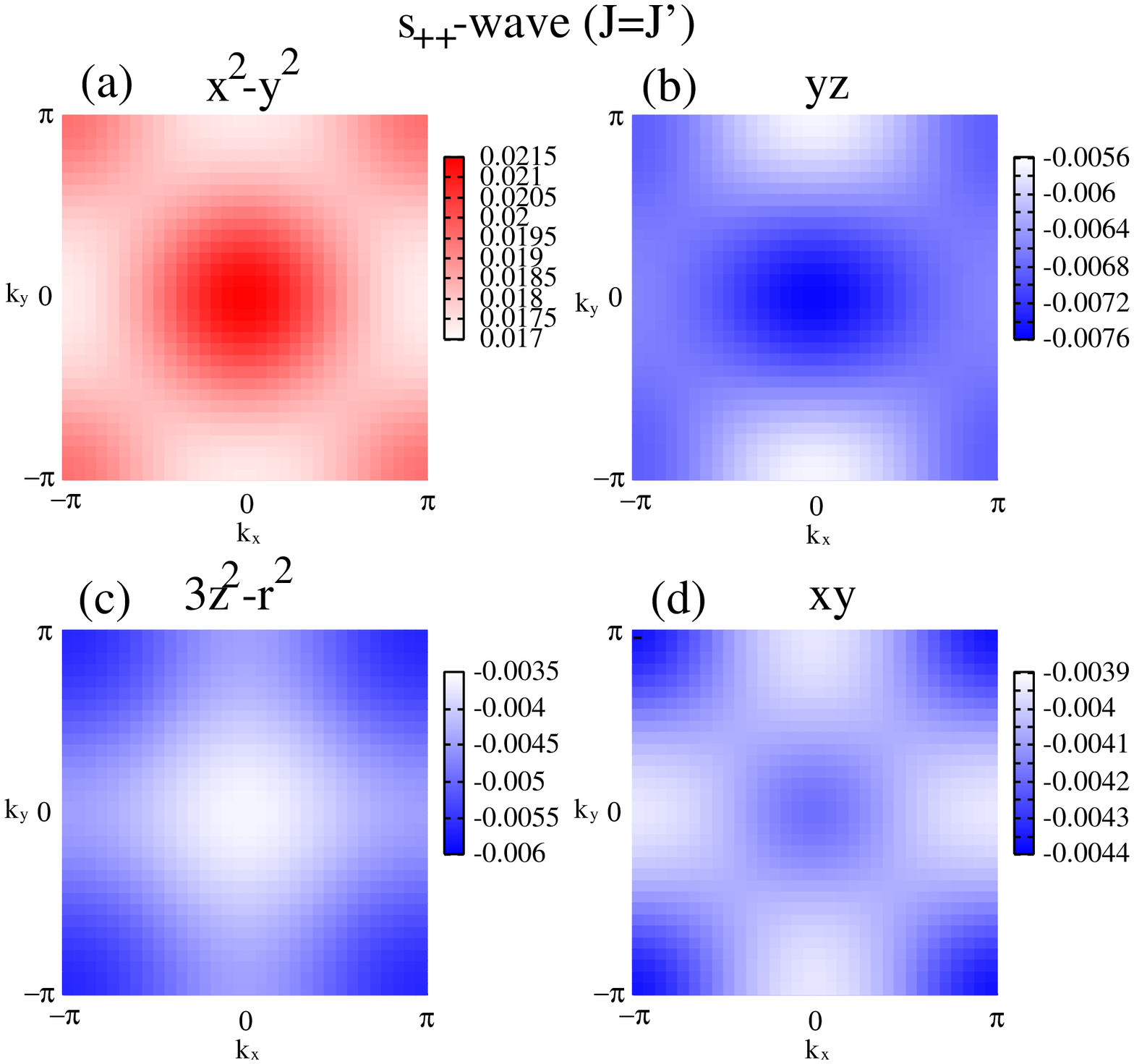}
\includegraphics[height=80.0mm,angle=-90]{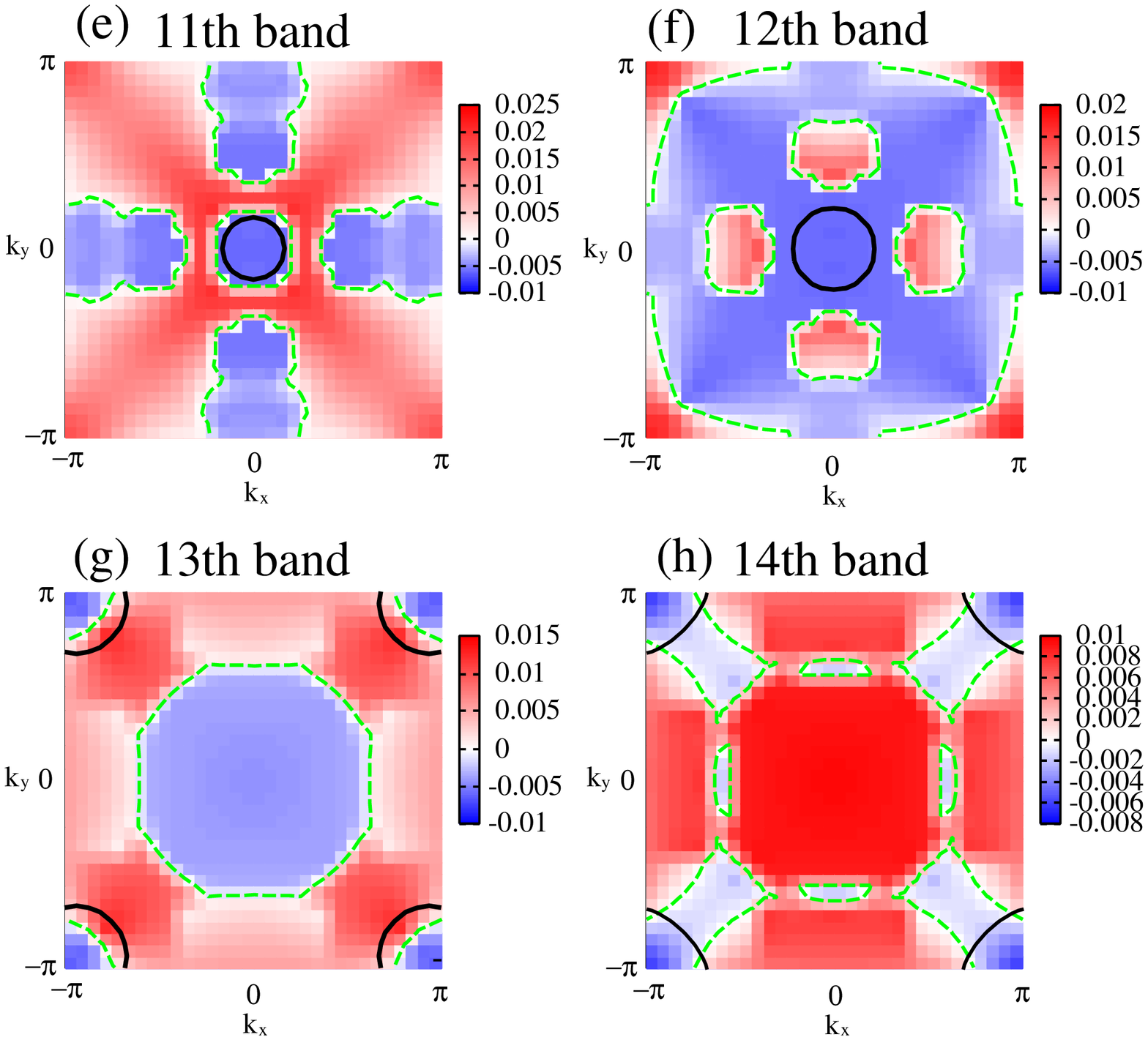}
\caption{(Color online) (a), (b), (c), (d) The diagonal components of the gap function
 $\hat{\Delta} (\mathbf{k})$ in the orbital representation and (e), (f),
 (g), (h) those in
 the band representation for $U=0.4$, $U'=1.15$ and $J=J'=0.1\mathrm{eV}$. The solid
 and dashed lines represent the Fermi
 surfaces and the nodes of the gap function, respectively. \label{fig_gap_2}}
\end{center}
\end{figure}
\begin{figure}[t]
\begin{center}
\includegraphics[height=83.0mm,angle=-90]{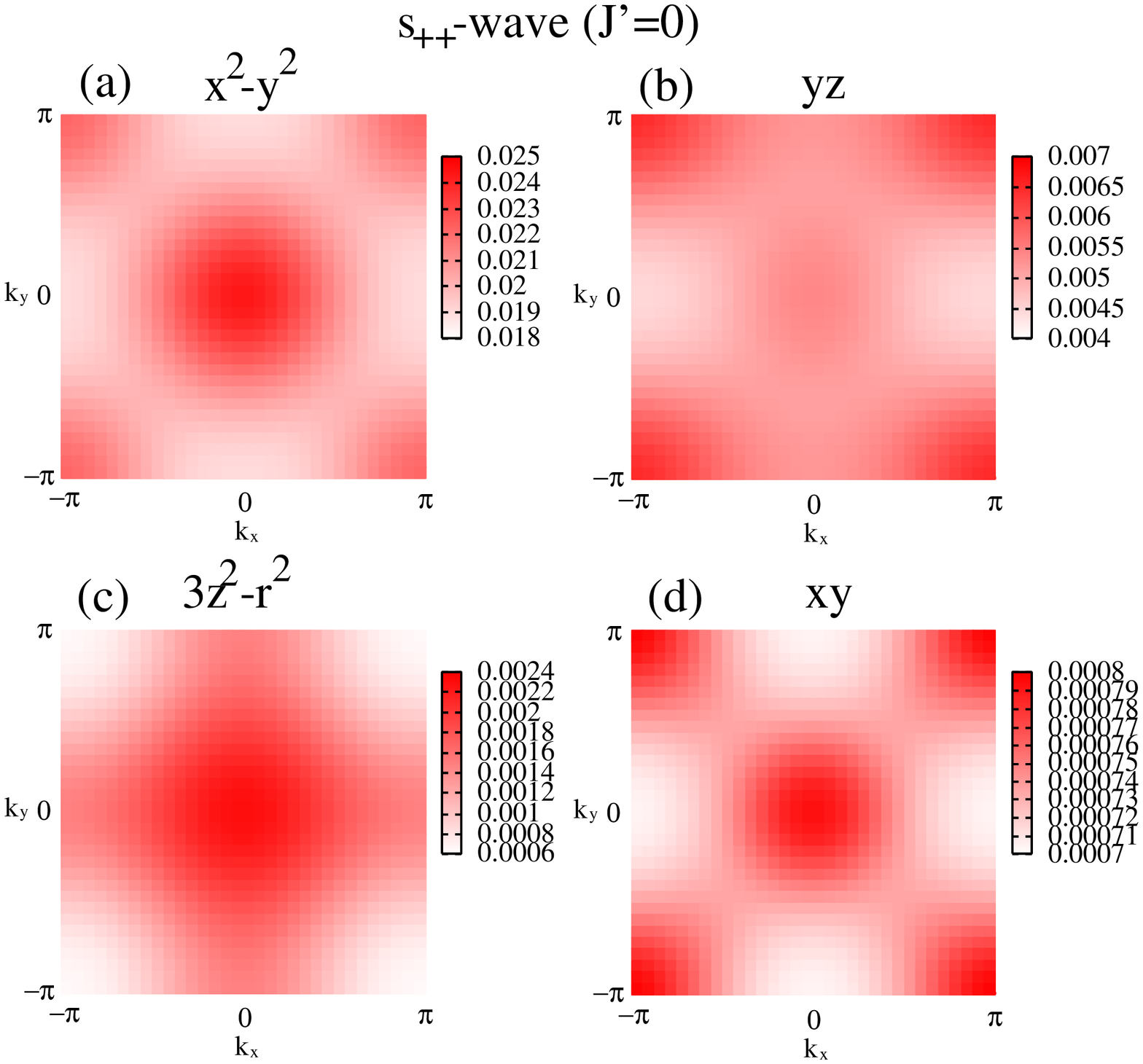}
\includegraphics[height=80.0mm,angle=-90]{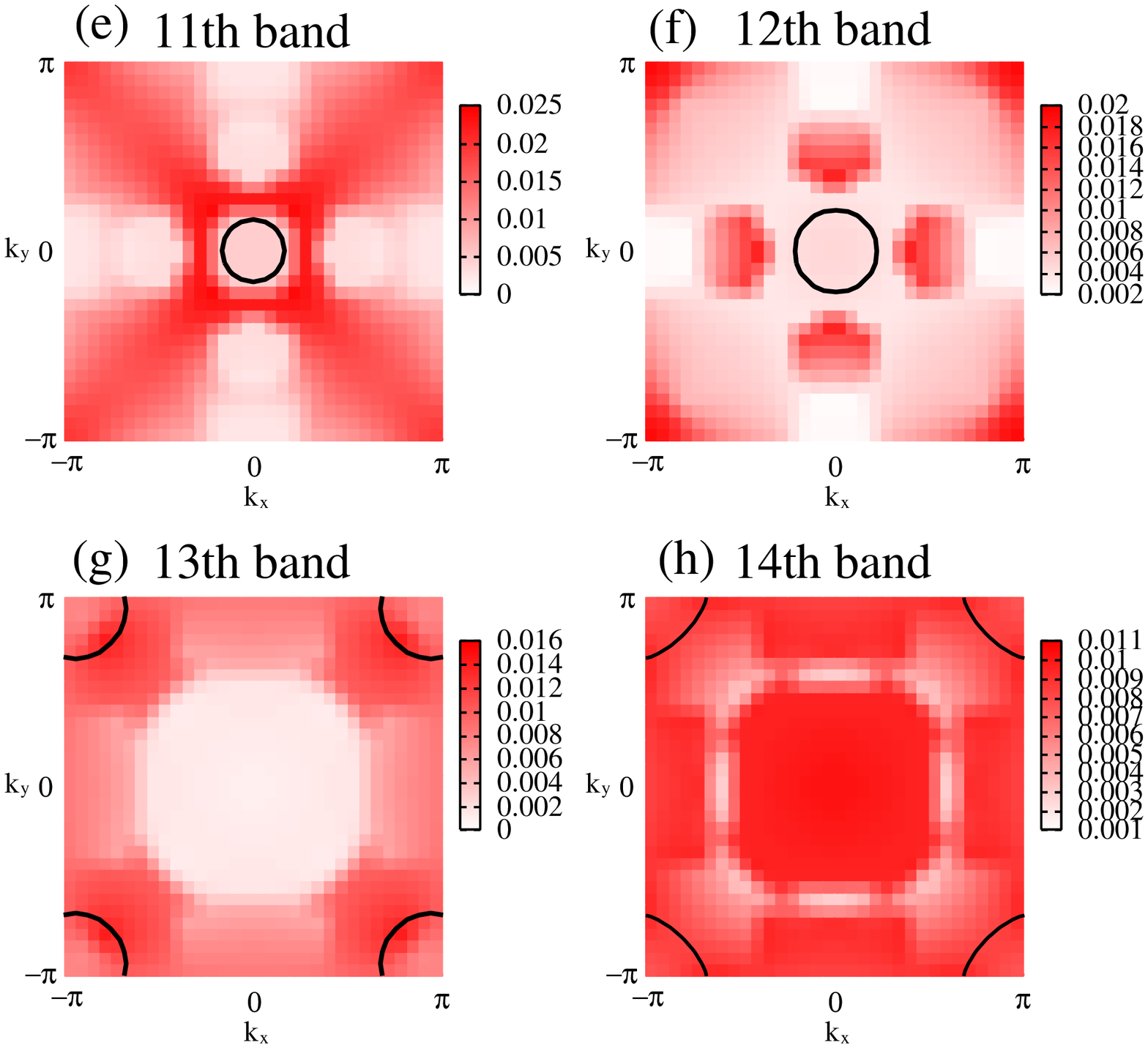}
\caption{(Color online) (a), (b), (c), (d) The diagonal components of the gap function
 $\hat{\Delta} (\mathbf{k})$ in the orbital representation and (e), (f),
 (g), (h) those in
 the band representation for $U=0.4$, $U'=1.18$ and $J=0.1$, $J'=0\mathrm{eV}$. The solid
 and dashed lines represent the Fermi
 surfaces and the nodes of the gap function, respectively. \label{fig_gap_3}}
\end{center}
\end{figure}
\subsection{Gap Functions}
First, we discuss the gap functions in the case with $U>U'$.
  Fig. \ref{fig_gap_1} shows the diagonal components of the gap function
  $\hat{\Delta}(\mathbf{k})$  for $U=1.71$, $U'=1.4$
  $J=J'=0.1$. Figs.  \ref{fig_gap_1} (a)-(d) show the gap
  functions in the orbital representation and Figs. \ref{fig_gap_1} (e)-(h) show those in the
  band representation. We note that the energy bands are numbered as
   descending energy.  It is found that the gap function has the
   $s$-wave symmetry and the
  most dominant component is the $d_{x^2-y^2}$ diagonal component.  We
  find that the gap functions in the band representation have different signs 
between the electron pockets and the hole pockets without any nodes on
the Fermi surfaces ($s_{\pm}$-wave
symmetry)\cite{mazin,kuroki_1,kuroki_2,kuroki_3,nomura_2,wang_1,ikeda,graser,yao,stanescu,cvetkovic}.
It is noted that the
  diagonal components of the gap function in the orbital representation, except
  for the $d_{xy}$ component, also change those signs in $\mathbf{k}$ space.
The absolute values of the
gap functions on the Fermi surfaces are almost isotropic but largely depend
on the energy bands; those on the electron pockets of the 13th and 14th
bands are twice or more larger than those on the hole pockets of the 11th
and 12th bands. This is because the $d_{x^2-y^2}$ component, which has dominant
contribution in $\hat{\chi}^s(\mathbf{q})$ as shown in Fig. \ref{chi} (a), for the 13th and 14th bands is
larger than that for the 11th and 12th bands. We note that the 10th band
(hole band) with the largest $d_{x^2-y^2}$ component has the largest absolute value
of the gap function, although the Fermi level is just above the 10th band and
does not cross it for $x=0.1$.


Next, we discuss the gap functions in the case with $U<U'$.
  Fig. \ref{fig_gap_2} shows the diagonal components of the gap function
  $\hat{\Delta}(\mathbf{k})$ for $U=0.4$, $U'=1.15$ 
  $J=J'=0.1$.  Figs. \ref{fig_gap_2} (a)-(d) shows
  the gap 
   functions in the orbital representation and Figs. \ref{fig_gap_2} (e)-(h) show those in the
  band representation. The diagonal components of the gap function in
  the orbital representation  have no sign change in the $\mathbf{k}$
  space due to the diagonal  components of $\hat{V} (\mathbf {\mathbf{q}})<0$ as
  shown in Fig. \ref{chi-2} (c).  We call this $s$-wave state as the $s_{++}$-wave state. The gap
  function in the band representation,  however, has sign change between
  the Fermi surfaces and line nodes on 
  the 14th band Fermi surface. These facts  reflect that the sign change of the gap
  function in the orbital representation  between the $d_{x^2-y^2}$
  diagonal component and the other 
  orbital diagonal components. The 11th band and 12th band Fermi
  surface has mainly $d_{yz}$ and $d_{zx}$ orbital  character, while, the
  13th band Fermi surface has mainly $d_{x^2-y^2}$ orbital 
  character. Therefore, the gap function has different sign  between the
  hole pockets and the 13th band electron pocket. The 14th  band electron
  pocket has mainly $d_{yz}$ and $d_{zx}$ orbital character  away from
  the Brillouin zone boundary, while $d_{x^2-y^2}$ orbital character on
  the 14th band electron pocket
  is comparable with $d_{yz}$ and $d_{zx}$ one near the Brillouin zone boundary. Thus, the gap
  function on the 14th band electron  pockets
  has plus sign near the zone boundary and minus sign away from the zone
  boundary. By the simple mean field
  analysis of the pair transfer term of the interacting part of the
  Hamiltonian eq (\ref{eq_H_int}),
 \begin{eqnarray*}
\frac{J'}{2}\sum_{i,\ell\neq\bar{\ell},\sigma\neq\bar{\sigma}}
\langle d^{\dag}_{i\ell\sigma}d^{\dag}_{i\ell\bar{\sigma}}\rangle
\langle d_{i\bar{\ell}\bar{\sigma}}d_{i\bar{\ell}\sigma}\rangle
\propto\sum_{\mathbf{k},\ell\neq\bar{\ell}}
\Delta_{\ell\ell}^{AA}(\mathbf{k})\Delta_{\bar{\ell}\bar{\ell}}^{AA}(\mathbf{k}).
\end{eqnarray*}
It is shown that the pair transfer $J'>0$ favors the sign change between
 the diagonal components of the gap function in the orbital representation.

  In fact, we also examine the case with $J'=0$ and we find that the
  $s_{++}$-wave state without sing change between the orbitals is
  realized for $U<U'$. We show the gap function for $U=0.4$, $U'=1.18$, $J=0.1$,
  $J'=0$ in Fig. \ref{fig_gap_3}. It is found that the all diagonal components of the gap
  function in the orbital representation have the same sign and those in the band
  representation have no sign change between all Fermi
  surfaces. Therefore, it is considered that the sign change of the gap function between the
  the $d_{x^2-y^2}$ diagonal component and the
  others is due to the pair transfer $J'$\cite{jkondo,yamaji}.

 It is helpful for understanding the difference between the
  $s_{\pm}$-wave state and the $s_{++}$-wave state in more detail to consider the gap function in the real
  space.  For $s_{\pm}$-wave state, the on-site pairing is comparable with the nearest neighbor 
   and/or the next nearest neighbor one. On the other hand, for
  the $s_{++}$-wave state, the on-site pairing is dominant and the
  off-site pairings are negligibly small as
  compared to the on-site pairing.

Here we discuss the reason why the on-site  part of the gap function  for $s_{\pm}$-wave state is large (especially in the $d_{x^2-y^2}$ diagonal component) even though the most dominant component of the effective interaction is always repulsive in $\mathbf{q}$ space (see Fig. \ref{chi} (c)), and then the on-site effective interaction is repulsive. 
When we perform the Fourier transformation of the gap equation eq. (\ref{eq_gap}), the on-site part of the left hand side is proportional to the on-site gap function, while that of the right hand side is given by the product  of the on-site effective interaction ($\times (-1)$) and the on-site anomalous Green's function which is proportional to the $\mathbf{q}$ summation of the gap-function times the single-particle spectral weight times the thermal factor. In the case with the $d_{x^2-y^2}$ diagonal component, the on-site gap function is negative as the negative contribution of the gap function in $\mathbf{q}$ space is much larger than the positive one as shown in Fig. \ref{fig_gap_1} (a). On the other hand, the on-site anomalous Green's function becomes positive as the single-particle spectral weight of the $d_{x^2-y^2}$ hole band is very large around the $\Gamma$ point where the gap function is positive as compared to that of the electron band around the $M$-point where the gap function is negative. Then, the gap equation can be satisfied with the large value of the on-site gap function against the repulsive on-site effective interaction. 

When the doping $x$ increases, the Fermi level rises apart from the $d_{x^2-y^2}$ hole band, and then the effect of the hole band decreases resulting in the decrease in the on-site gap function as well as the decrease in the superconducting transition temperature (not shown). Such doping dependence of the on-site gap function has recently been observed in the 5-band Hubbard model\cite{kariyado}. 
On the contrary, in the $s_{++}$-wave state, the on-site pairing is always dominant almost independent of the doping $x$. 

\subsection{Phase Diagram}
The phase diagram on $U'$-$U$ plane for $J=J'=0.1$ is shown
in Fig. \ref{phasediagram}, where the magnetic and charge-orbital
instability is determined by 
$\mathrm{det}(\hat{1}-\hat{\chi}^{(0)}(\mathbf{q})\hat{S})=0$ and 
$\mathrm{det}(\hat{1}-\hat{\chi}^{(0)}(\mathbf{q})\hat{C})=0$,  
respectively and the superconducting 
instability is determined by $\lambda=1$ as mentioned before.  
 The ISDW with $\mathbf{q}\sim(\pi,\pi)$ appears in the large $U$ 
region, while, the ferro-orbital order appears for $U<U'$\cite{adachi} [see also 
Figs. \ref{chi} (a), (b) and Figs \ref{chi-2} (a), (b)]. It is noted
that on the phase boundary where the charge-orbital instability takes place, 
the longitudinal orbital susceptibility
$(\hat{\chi}^c(\mathbf{q}))^{\alpha\beta}_{\ell\ell,\ell'\ell'}$ diverges, while, the charge susceptibility $\sum_{\ell,\ell',\alpha,\beta}
(\hat{\chi}^c(\mathbf{q}))^{\alpha\beta}_{\ell\ell,\ell'\ell'}$ dose not
. The $s_{\pm}$-wave
pairing is realized near the ISDW due to the spin fluctuations, while,
the $s_{++}$-wave pairing is 
realized near the ferro-orbital ordered phase due to the charge-orbital 
fluctuations, where we regard the superconducting states as the 
$s_{++}$-wave states if $d_{yz}$, $d_{zx}$ and $d_{x^2-y^2}$ diagonal 
components of the gap function have no sign change in $\mathbf{k}$ space
and as the $s_{\pm}$-wave states if not. The way to determine whether
the superconducting state is the  $s_{\pm}$-wave state or the
$s_{++}$-wave state is not unique. This is because the $s_{\pm}$-wave
and the $s_{++}$-wave state are same symmetry (A$_{1g}$) and the change
between $s_{++}$-wave and the $s_{\pm}$-wave
state is crossover. In fact, as $U$ increases, the on-site
paring decreases, while, the off-site pairing increases continuously. At $U\sim1.25$, the nodes
appear around the $M$-point for the $d_{x^2-y^2}$ diagonal component and
those approaches the $\Gamma$-point as $U$ increases. It is noted that we also
obtain the phase diagram on $U'$-$U$ plane 
for $J=J'=0.25 $ and find that the phase diagram is essentially the same
as that for $J=J'=0.1$ except that the magnetic and the $s_{\pm}$-wave superconducting instabilities are
slightly enhanced by the larger value of the Hund's coupling $J$. 

\subsection{Effects of Electron-Phonon Coupling}
In this subsection, we discuss the effects of the electron-phonon
coupling. 
By performing the group theoretical analysis for LaFeAsO, it is found
that there are 14 kinds of
the optical phonon modes at the $\Gamma $ point:
2$A_{1g}$+2$B_{1g}$+4$E_g$+3$A_{2u}$+3$E_u$.
Here, we concentrate on the $A_{1g}$ mode in which La
and As ions oscillate along the c-axis. The $A_{1g}$ phonon dose not
break the symmetry of the orbital and the resulting electron-phonon coupling
matrix $\hat{g}$ is diagonal in the orbital representation. Within the RPA, the charge-orbital susceptibility $\hat{\chi}^c
(\mathbf{q})$ including the effects of both the electron-electron and the
electron-phonon coupling is obtained
by replacing $U$ with $U-2U_{ph}$ and $2U'-J$ with
$2(U'-U_{ph})-J$ in eqs. (\ref{eq_chic}) and (\ref{eq-U}), 
where $U_{ph}=2g^2/\omega_{A_{1g}}$, $\omega_{A_{1g}}$ is the frequency of the $A_{1g}$ phonon and we
neglect the orbital- and $\mathbf{q}$-dependence of the electron phonon interaction. It is found
that the inter-orbital direct term $U'$ which enhances the orbital fluctuations is harder to be reduced
by the electron-phonon coupling than the intra-orbital direct term
$U$. As a result, the orbital fluctuations are relatively enhanced by
the electron-phonon coupling as compared to the spin fluctuations. 
 
  \section{Summary and Discussion}
 In summary, we have investigated the pairing symmetry of the two-dimensional
 16-band $d$-$p$ model by using the
 RPA and have obtained the phase diagram including the magnetic and orbital 
 orders and the superconductivity.  
 For $U>U'$, the $s_{\pm}$-wave superconductivity is realized near the
 ISDW with $\mathbf{q}\sim(\pi, \pi)$ phase. On the other hand, for $U<U'$, the
 $s_{++}$-wave superconductivity appears near the ferro-orbital ordered phase. 
 The $s_{\pm}$-wave pairing  is 
 mediated by the spin fluctuations, while that the $s_{++}$-wave pairing is
 mediated by the orbital fluctuations.

 For $U>U'$, the gap function for the $s_{\pm}$-wave pairing changes 
 its sign between the hole pockets and the electron pockets
 and the most dominant contribution of the gap function is the 
 $d_{x^2-y^2}$ orbital diagonal component. This is qualitatively
 consistent with the results based on the 5-band Hubbard
 model\cite{kuroki_1,kuroki_2,kuroki_3,nomura_2,ikeda,wang_1}. 
 However, the $d_{x^2-y^2}$ diagonal component of the gap function 
 in our 16-band $d$-$p$ model have 
 much larger value than the other matrix elements in comparison with
 the results based on the 5-band Hubbard
 model\cite{kuroki_2,nomura_2}. This may be because the
 outer hole Fermi surface which has mainly $d_{yz}$ and $d_{zx}$ orbital 
 character obtained by the $d$-$p$ model is almost 
 circular, but that obtained by the 5-band Hubbard model is 
 diamond shape\cite{kuroki_1}.  Therefore, the nesting effect 
 which enhances the spin fluctuations  becomes weak in our $d$-$p$ model, 
 and the resulting components of 
 $\hat{\Delta} (\mathbf{k})$ related to $d_{yz}$, $d_{zx}$ orbitals 
 have smaller values. 
 
 For $U<U'$, the gap function in the orbital representation  
 for the $s_{++}$-wave pairing dose not change its sign in $\mathbf{k}$ space. 
 In other words, the on-site pairing is much larger than the off-site
 pairing in the real space.  This is similar to the conventional
  phonon-mediated superconductivity. However, the gap functions have different
  signs between orbitals in contrast to the conventional phonon-mediated
  superconductivity.  We have shown that this sign change of the gap functions
  between orbitals is due to the effect of the  pair transfer interaction 
  $J'$\cite{jkondo,yamaji}.
  It is noted that the $s_{++}$-wave state has been observed also in the
  one-dimensional 2-band Hubbard model in the same parameter region with 
  $U<U'$ \cite{sano}.

  It seems that the both $s_{\pm}$-wave and the $s_{++}$-wave states 
  with full superconducting gaps are 
  consistent with various experiments such as, 
  the NMR relaxation rate, the Knight shift, the ARPES, the magnetic 
  penetration depth measurements, although the sign of the gap function has
  not been directly observed there. However, according to
 the recent theoretical studies of the nonmagnetic impurity
 effects\cite{onari}, Anderson's
 theorem is violated for the $s_{\pm}$-wave superconductivity in contrast
 to the experimental results of very weak $T_c$-suppression in Fe site
 substitution\cite{kawabata_1} and neutron irradiation\cite{karkin}.
 Since it can be considered that the impurity potential by the Fe-site
 substitution is 
 diagonal and local in the orbital basis according to the first  principle
 calculation\cite{kemper}, it is expected that the
 $s_{++}$-wave state observed in the present study is  more
 robust against the nonmagnetic impurity effects than the $s_{\pm}$-wave
 state.

  In addition to the Coulomb interaction,  we have also discussed 
  the effects of the coupling $g$ between the electron and the $A_{1g}$ phonon 
  within the RPA. It has been found that the $s_{++}$-wave pairing realized 
  in the unrealistic parameter region with $U<U'$ for $g=0$ is enhanced due to 
  the effect of $g$ and can be expanded over 
  the realistic parameter region with $U>U'$ for a realistic value of $g$. 
  In the first principle calculations for iron-based superconductors 
  in conjunction with the Migdal-Eliashberg theory, 
  the electron-phonon coupling is found to be too small to obtain 
  high $T_c$ observed in experiments\cite{boeri}. 
  The effect of the Coulomb interaction, however, has not been 
  discussed there.  
  In the present study, the cooperative effect of the Coulomb 
  interaction and the electron-phonon coupling is crucial for 
  the enhancement of the orbital fluctuations which induce 
  the $s_{++}$-wave superconductivity. 
  Recently, the large isotope effects on the transition temperatures 
  for both the SDW and the superconductivity have been observed\cite{liu_2}. 
  This experimental result implies that not only the Coulomb interaction 
  but also the electron-phonon coupling plays crucial effects 
  on the electronic states for iron-based superconductors.

  In early theoretical studies for the copper oxide superconductors, the effect 
  of the Coulomb interaction between the $d$ and $p$ electrons $U_{pd}$ 
  was studied by several authors\cite{littlewood_2,hirashima}.
  According to the RPA study based on the $d$-$p$ model with 
  the single $d_{x^2-y^2}$ orbital, 
  $U_{pd}$ enhances the charge fluctuations with $\mathbf{q}=(0,0)$ and the
 $s$-wave superconductivity is realized due to the effect of  charge 
 fluctuations\cite{littlewood_2}. In addition, the $1/N$-expansion approaches 
  ($N$ is the spin-orbital degeneracy) revealed that 
  the strong correlation effect enhances the charge fluctuations together
 with the $s$-wave superconductivity\cite{hirashima}. Therefore, it is expected
 that, in the present $d$-$p$ model with multi $d$ orbitals, 
 $U_{pd}$ enhances  the charge-orbital fluctuations which induce  
 the $s_{++}$-wave superconductivity.
 The explicit calculations based on the $d$-$p$ 
 model including not only the on-site Coulomb interaction but also the 
 inter-site Coulomb interaction $U_{pd}$ together with the electron-phonon 
 coupling $g$ are now under way. 

\begin{acknowledgments}
The authors thank M. Sato, H. Kontani, S. Onari, T. Nomura, H. Ikeda, K. Kuroki and Y. Yanase for useful
comments and discussions. This work was partially supported by the
Grant-in-Aid for Scientific Research from the Ministry of Education,
Culture, Sports, Science and Technology. 
\end{acknowledgments}


\begin{thebibliography}{99}
\bibitem{notation}  The spin and charge-orbital susceptibilities are the 
      particle-hole channel, while the superconducting susceptibility is the 
      particle-particle channel. Then, there is a rotation of the orbital 
      indices of the effective pairing interaction $V$ in the gap equation 
      eq.(\ref{gapeq}) with respect to 
      those of the spin and charge-orbital susceptibilities 
      in eqs.(\ref{eq_chis})-(\ref{eq_chi0}). 
      This is more explicitly represented in ref. 31 with a different notation.
\end{thebibliography}
\end{document}